\documentstyle[12pt]{article}
\def \beq{\begin{equation}}
\def \eeq{\end{equation}}
\def \pa{\partial}
\newcommand{\n}{
\newpage
\noindent
}
\begin{document}
\n
\pagestyle{empty}

\begin{center}
{\large \bf  Macroscopic   Einstein - Maxwell  equations   for   a   system   of
interacting particles to  second-order  accuracy  in  the interaction
constan.}

\bigskip
     A. V. Zakharov

\medskip
{\it Dept.Gen.Rel.\& Grav.\\
     Kazan State University,\\
     Kremljevskaja Str. 18\\
     420008 Kazan, Russia\\
     e-mail: Alexei.Zakharov@ksu.ru}
\end{center}

\newpage

\begin{center}
{\large \bf  Macroscopic   Einstein - Maxwell  equations   for   a   system   of
interacting particles to  second-order  accuracy  in  the interaction
constant.}

\bigskip

     A. V. Zakharov

\bigskip

{\bf Abstract.}

\end{center}

\bigskip

In this paper the macroscopic Einstein and Maxwell equations for
system, in which  the electromagnetic interactions
are dominating (for instance, the cosmological
plasma before the moment of recombination), are derived.

Ensemble averaging of  the  microscopic  Einstein - Maxwell equations
and  the
Liouville equations  for the random functions leads to a closed system
of macroscopic Einstein - Maxwell equations and  kinetic  equations  for
one-particle distribution  functions.
The  macroscopic Einstein equations
for a relativistic plasma differ  from
the classical  Einstein  equations  in  that  their  left-hand sides
contain additional terms due to particle interaction.  The terms  are
traceless tensors   with   zero   divergence.  An  explicit  covariant
expression for these terms is given in the form of  momentum-space
integrals of  expressions  depending  on  one-particles distribution
functions of the interacting particles of the medium.

The additional terms in the  left-hand  side  of  the  macroscopic
Einstein equations for a relativistic plasma has murch in common with the
additional terms in the  left-hand  side  of  the  macroscopic
Einstein equations  for  a  system  of self - gravitating
particles (refer to.  [1], [2]).

     The macroscopic Maxwell equations alsow differ from the classical
macroscopic Maxwell  equations  in  that their left-hand sides contain
additional terms due to particle interaction as well the effects  of
general relativity.

\newpage
\pagestyle{plain}

\setcounter{page}{1}

\begin{center}
{\large \bf  Macroscopic   Einstein - Maxwell  equations   for   a   system   of
interacting particles to  second-order  accuracy  in  the interaction
constant.}

\bigskip

     A. V. Zakharov
\end{center}
\bigskip

{\bf 1. Introduction.}

\bigskip

The idea of macroscopic gravity can be considered as an  extension  of
Lorentz' idea (refer to. [3]),  formulated  first for electrodynamics,
about the
existence of two levels,  microscopic and macroscopic, of understanding
classical physical    phenomena.   Lorentz   showed   that   Maxwell's
electrodynamics is a macroscopic theory of electromagnetism,  and  the
Maxwell equations  could be  derived from a system of microscopic field
equations called now the Maxwell  -  Lorentz  ones,  by  infinitisimal
space - time regions averaging  (refer to. [3], [4]).

As it is known the macroscopic Maxwell equation for continuous media  can
be alsow obtained  from  the  microscopic  Maxwell  equations  by  ensemble
averaging the latter (refer to.[5]).

The Einstein equations,  whose right-hand side contain the
energy-momentum tensor  of  matter,  are  phenomenological  equations.
It is
natural to  suppose   that   the   Einstein   equations   (or   their
generalizations) for  continuous  media  can also be obtained from the
microscopic Einstein equations, i.e., Einstein equations whose
right-hand sides  contain  the  sum  of  the  energy  -  momentum tensors of
individual particles.  However,  due to the nonlinearity of the
left-hand side  of  Einstein equations,  the averaging of the
microscopic Einstein
equations is more complicated than one of the  microscopic  Maxwell
equations (refer to. [6] -  [10]).

In (refer to.[1])  a  method  is  developed  for  the  ensemble
averaging of the
microscopic Einstein equations for interacting particles.
( We use  the  ensemble  averaging
procedure introduced   by  Klimontovich  (refer to. [11,  12])
to  derive  the
relativistic kinetic equation for a plasma.  The  same  procedure  was
used by  the  present  author  in (refer to. [13,  14])
to derive a relativistic
kinetic equation for a system of gravitationally and electromagnetical
interacting particles  in  General  Relativity  accurate to within the
second order for the interaction constant.)
This results to macroscopic Einstein equations  for  continuous  media
that  are accurate to second-order terms
in the interaction constant.
The  macroscopic Einstein equations
for a system of interacting particles  differ  from
the classical  Einstein  equations  in  that  their  left-hand sides
contain additional  components  due  to  particle   interaction.   The
components are
expressed in terms of the two-particle correlation function  of
the particles.

     In (refer to. [2])  we  got   covariant expressions for
additional components
for the system of self-gravitating particles.
The components are
traceless tensors   with   zero   divergence.
The expressions were obtained in
the form of momentum-space integrals of expressions depending on
one-particle distribution function of the  gravitationally  interacting
particles of the medium.
The given expressions are proportional to the
cube of the Einstein constant and the square of  the  particle  number
density. The  latter relationship    implies   that  interaction  effects
  manifest
themselves in systems of very high density (the Universe in the  early
stages of   its   evolution,  dense  objects  close  to  gravitational
collapse, etc.)

The present paper is a direct continuation of  earlier papers
(refer to.[1], [2]),
devoted to  the derivation of the macroscopic Einstein equations for a
system of self-gravitating particles to within terms of second order
in the interaction constant.

The objective of this paper is to obtain the macroscopic
Einstein equations
for system, in which  electromagnetic
interactions (for instance, cosmological plasma before a moment
of recombination,) are dominating.

The  macroscopic Einstein equations
for relativistic plasma  differ  from
the classical  Einstein  equations  in  that  their  left-hand side
contains additional terms due to particle interaction.
The terms  are
traceless tensors   with   zero   divergence.  An  explicit  covariant
expression for these terms is given in the form of  momentum-space
integrals of  expressions  depending  on  one-particles distribution
functions of the interacting particles of the medium.

The additional terms in the  left-hand  side  of  the  macroscopic
Einstein equations  for  a  relativistic plasma has march in common with
the additional terms in the  left-hand  side  of  the  macroscopic
Einstein equations  for  a  system  of self - gravitating
particles (refer to.  [1], [2]).

     The macroscopic Maxwell equations alsow differ from the classical
macroscopic Maxwell  equations  for their left-hand sides contain
additional terms due to particle interaction as well the  effects  of
general relativity.

{\bf 2.   Microscopic   Einstein - Maxwell  equations}
\bigskip

The method of deriving the macroscopic Einstein equations is discussed
in (refer  to.[1]).  The  notation  we  use  here are the same that in
(refer to.[1]).

     Briefly, the method we used to obtain  the  macroscopic  Einstein  -
Maxwell equations is the following.

     We start from the microscopic Einstein and Maxwell  equations

\beq \label {1}
 \tilde G^{ij}=\chi\tilde T^{ij}_{(m)} + \chi\tilde T^{ij}_{(el)},
\eeq

\beq \label {2}
\tilde \nabla_{k} \tilde F^{ik} = - \frac {4\pi}{c} \tilde J^i.
     \eeq

Here $ \tilde G^{ij}$ is the Einstein tensor  in  a  Riemannian  space
with metric  $  \tilde  g_{ij}$  ,  $  \chi  =  {8\pi  k}/{c^{4}}$  is
Einstein's constant (where $k$ is the gravitational constant,  $c$  is
the velocity   of   light),   $\tilde   T^{ij}$   is  the  microscopic
energy-momentum tensor of particles,  $\tilde F^{ik}$ is the Maxwell's
tensor,  $\tilde J^i$ is the microscopic current vector of particles,
$\tilde T^{ij}_{(el)}$   is   the   energy   -   momentum   tensor  of
electromagnetical field.  Raising   and   lowering   of   indexes   is
accomplishment with the metric $\tilde g_{ij}$, $\tilde \nabla_{k}$ is
a covariant derivative in a  Riemannian  space with metric
$  \tilde  g_{ij}$.

     The tensor $\tilde  T^{ij}_{(el)}$ have the form
\beq \label {3}
\tilde  T^{ij}_{(el)} = \frac {1}{4\pi}\left(- \tilde F^i_{.l}
\tilde F^{lj}  + \frac {1}{4} \tilde g^{ij}\tilde F_{lm} \tilde F^{lm}
\right).
     \eeq

     The tensors $\tilde  T^{ij}_{(m)}$ and $\tilde J^{i}$ has the form
     \beq \label {4}
 \tilde T^{ij}_{(m)}  = \sum_a m_a c^2\int\frac{d^4\tilde p_a}{\sqrt {-\tilde g}}
\tilde u_{a}^{i}\tilde u_{a}^{j} {\tilde N}_a(q^i,\tilde p_i) ,
     \eeq
     \beq \label {5}
     {\tilde J}^i = \sum_b e_bc \int \frac {d^4 \tilde  p}
{\sqrt  {-\tilde  g}}
{\tilde u}^i_b {\tilde N}_b(q, \tilde p).
     \eeq

     Here $e_a$  is  the charge of particles of species "a",  $m_a$ is
their mass, $\tilde g$  is the determinant of $\tilde  g_{ij}$,  $\tilde
p^{i}_{a}$ is the momentum of particles of spesies "a",
$\tilde u^{i}_{a} = {\tilde p^{i}_{a}}/{\sqrt {   \tilde    g_{kj}\tilde
p^{k}_{a}\tilde p^{j}_{a}}}$,
      $$\frac  {d^4  \tilde p}{\sqrt  {-\tilde g}}  $$
     - is the invariant volume element in momentum spase [7].

$N_{a}(q^i,\tilde p^{i}_{a})$  -  is  the Klimontovich random function
[11]:
     \beq \label {6}
     \tilde N_a(q^i,\tilde p_j) = \sum^{n_a}_{i=1}\int d\tilde s
\delta^4(q^i -    q^{i}_{(l)})\delta^4(\tilde     p_j     -     \tilde
p^{(l)}_{j}(\tilde s)).
\eeq
    Nere  $n_a$ is the number of particles belonging to species "a",
$\tilde s$ is the canonical parameter along the particle trajectories:
$d\tilde s = \sqrt {g_{ij}dq^idq^j}$; $q^{i}_{(l)}$  and
$\tilde p^{(l)}_{j}$  are  the  coordinates  and  momentum of the l-th
particle of spesies "a",  which are found by solving the equations  of
motion:
\beq \label {7}
     \frac {dq^i_{(l)}}{d \tilde s} = \frac {\tilde  p^i_{(l)}}{m_ac},
\qquad \frac  {d \tilde p_i^{(l)}}{d \tilde s} = \frac {1}{m_ac}
\tilde \Gamma_{j,ik} \tilde p^j_{(l)} \tilde p^k_{(l)} + \frac {e_a}{c}\tilde
F_{ik} \tilde p^k_{(l)}.
     \eeq

Here $\tilde  \Gamma_{j,ik}$  is  the  Christoffel symbol of the first
kind given by the metric $\tilde g_{ij}$.

     In view of Eqs.  (\ref {7}) the random function (\ref {6})  obeys
the equation
     \beq \label{8}
     \tilde p^i\frac{\pa   {\tilde   N_a}}{\pa   {q^i}}    +    \tilde
\Gamma_{j,ik}\tilde p^j\tilde   p^k\frac{\pa{\tilde   N_a}}{\pa{\tilde
p_i}} + \frac {e_a}{c} \tilde F_{ik}  \tilde  p^k  \frac  {\pa  \tilde
N_a}{\pa \tilde p_i} = 0.
     \eeq

     Next we write the metric $\tilde   g_{ij}$ as
     \beq \label {9}
     \tilde g_{ij} = g_{ij} + h_{ij},
     \eeq
     and $\tilde F_{ik}$  as
     \beq \label {10}
     \tilde F_{ik} = F_{ik} + \omega_{ik},
     \eeq

    Here $g_{ij} = \langle \tilde g_{ij} \rangle  $  is  the  ensemble
averege of the metric  $\tilde  g_{ij} $ [1, 2],
$\langle \tilde F_{ik} \rangle$ is the ensemble average of $\tilde F_{ik}$.
Note that $\langle h_{ij}\rangle \equiv 0$ and
$\langle \omega_{ik} \rangle \equiv 0$.

Parallel with the momenta
$\tilde  p^i_{(l)} = m_ac dq^i_{(l)}/d \tilde s$
we use the momenta $p^i$ measured in the metric $g_{ij}$:
     \beq \label{11}
     p_{(l)}^i = \alpha^{-1}(q,p)\tilde p^i_{(l)}, \qquad \alpha(q,p) =
     ds/d\tilde s   =   (g_{ij}p^ip^j)^{1/2}(\tilde  g_{lk}\tilde  p^l
\tilde p^k)^{-1/2}.
     \eeq
     Here $s$ is the canonical parametr introduced by $g_{ij}$.

The transformation from $\tilde p_i$ to $p_i$ is given by
     \beq \label{12}
     \tilde p_j   =   \tilde   g_{jk}   \tilde  p^k  =  \alpha  \tilde
g_{jk}g^{ki}p_i.
     \eeq
    THe Jacobian of transformation (\ref {12}), is (see [14]):
     \beq \label{13}
     \mid \frac{\pa {\tilde p}_i}{\pa p_j} \mid = \alpha^4 \frac
{\tilde g}{g},
     \eeq
     where $g$ is the determinant of $g_{ij}$.

     Now we  introduce  the function $ N_a(q^i,\tilde p_j)$ defined in
the eight - dimensional phase space with coordinates $(q,p)$ as
     \beq \label {14}
      N_a(q, p) = \sum^{n_a}_{l=1}\int d s
\delta^4(q^i -    q^{i}_{(l)}(s))
\delta^4(p_j -  p^{(l)}_{j}(s)),
\eeq
  where $q^i_{(l)}$ и $p_j^{(l)}$ are found  by  solving  equations
obtained from  (\ref   {7})  with the transformation  (\ref   {12})
taken into account ($p^i=g^{ij}p_j$).

     Note that the functions $\tilde N_a$ and $N_a$ are related in the
folloing manner:
     \beq \label {15}
     \tilde N_a(q,\tilde p) = \frac {g}{\tilde g \alpha^5} N_a(q,p).
     \eeq

Equation for $N_a(q,p)$ can be obtained directly from (\ref {8})
by replaicing the variables   (\ref {12}) and (\ref {15}):
    $$ p^i\frac{\pa   {  N_a}}{\pa   {q^i}}    +
\Gamma_{j,ik} p^j  p^k\frac{\pa  N_a}{\pa p_i} + \frac {e_a}{c} F_{ik}
p^k \frac {\pa N_a}{\pa p_j} = $$
     \beq \label{16}
= \frac  {\pa}{\pa
p_i}\Bigl[ \left( \Omega^m_{jk}\Delta_{mi}p^j  p^k  -  \frac  {e_a}{c}
\psi^l_{.k} \Delta_{lj} p^k \right) N_a \Bigr].
     \eeq

     Here
     \beq \label{17}
     \Delta_{ki} = g_{ki} - u_k u_i;  \qquad u^k =  \frac  {p^k}{\sqrt
{p^l p_l}};  \qquad \Omega^m_{kj} =  \tilde \Gamma^m_{kj} -
\Gamma^m_{kj} -
     \eeq
     - is the difference of the Christoffel symbols of second kind for
the metrics $\tilde g_{ij}$  and $g_{ij}$,
     \beq \label{18}
     \psi^l_{.k} =  \frac  {1}{\alpha  (q,p)}  \tilde   F^l_{.k}   -
F^l_{.k} =  \frac  {1}{\alpha  (q,  p)}  \tilde g^{lm} \tilde F_{mk} -
g^{lm} F_{mk}.
     \eeq
If in (\ref {4}) and (\ref {5}) we turn to the variables $p_i$ and
$N_a$ we get
     \beq \label {19}
   \tilde T^{ij}=  \sum_{a} m_a с^2\int   \frac{d^4p_a}{\sqrt{(-g)}}
\alpha(q,p)\sqrt{\frac{g}{\tilde g}} u_a^i u_a^j N_{a}(q,p_a),
     \eeq
     \beq \label {20}
     {\tilde J}^i = \sum_b e_bc \int \frac {d^4  p_b}
{\sqrt  {-g}} \sqrt {\frac {g}{\tilde g}}
{u}^i_b {N}_b(q, p_b).
     \eeq
where
      $ d^4 p /{\sqrt  {- g}}  $
     is the invariant volume element in the unperturbed momentum spase.

     For subsequent calculation is it convinient to write the Einstein
equations as
      $$ R_{ij} + \nabla_m \Omega^m_{ij} -
     \nabla_j \Omega^m_{im} +  \Omega^m_{mn} \Omega^n_{ij} -
     \Omega^m_{jn} \Omega^n_{im} = $$
     \beq \label {21}
     = \chi \sum_a m_a c^2 \int \frac {d^4 p_a}{\sqrt {-g}} \alpha
\sqrt  {\frac {g}{\tilde g}} \left( \tilde g_{ik} \tilde g_{jm} -
\frac {1}{2} \tilde g_{ij} \tilde g_{km}\right) u^k_a u^m_a N_a(q, p_a) +
\chi \tilde T_{ij}^{(el)}.
     \eeq

Here $R_{ij}$  is  the  Richi  tensor of the Riemannian spase with the
metric $g_{ij}$, $\nabla_m $ is the covariant derivative in this spase.

  With the  help  (\ref  {9}) and (\ref {10}) we can write the Maxwell
equations (\ref  {2}) to within the first-order terms in $h_{ij}$:

$$\nabla_k F^{ik} +    \nabla_k \left(h^i_m     F^{km}    -    h^k_m
F^{im}\right) + \frac {1}{2} F^{ik} \nabla_k h + \nabla_k \omega^{ik} + $$
     \beq \label {22}
     +     \nabla_k \left(h^i_m     \omega^{km}    -    h^k_m
\omega^{im}\right) + \frac {1}{2}  \omega^{ik} \nabla_k h  =  -
4\pi \sum_a e_a \int \frac {d^4 p_a}{\sqrt {-g}} u^i_{(a)}  \left(  1  -
\frac {1}{2} h \right) N_a(q, p_a).
     \eeq

In (\ref {21}), {\ref {22}) and below when raising and lowering the
indexes we use the averaged metric $g_{ij}$, $h=h^l_l$.

     Let's expand  the Einstein equations (\ref {21}) up to the secand
-order members in small $h_{ij}$ and $\omega_{ij}$:

      $$ R_{ij} + R^{(1)}_{ij} + R^{(2)}_{ij} + \cdots =
 \sum_a \chi m_a c^2 \int \frac {d^4 p_a}{\sqrt {-g}}
\left( \tilde g_{ik} \tilde g_{jm} -
\frac {1}{2} \tilde g_{ij} \tilde g_{km}\right) u^k_a u^m_a N_a(q, p_a) + $$
     \beq \label {23}
+ \sum_a \chi m_a c^2 \int \frac {d^4  p_a}{\sqrt  {-g}}  L^{(1)}_{ijkm}
(h) u^k_a   u^m_a  N_a  +  \cdots  +  \chi  T_{ij}^{(el)}  +
\chi  \left( T_{ij}^{(el)}\right)^{(1)} +
\chi  \left( T_{ij}^{(el)}\right)^{(2)}.
     \eeq

    Here $R_{ij}^{(1)} $ is the sum of all terms of expansion
$\tilde R_{ij}$ that are first-order in $h_{ij}$,
    $R_{ij}^{(2)} $ is the sum of all second-order terms in
$h_{ij}$, ets.;  $L^{(1)}_{ijkm}(h)$ is the sum of all first  -  order
terms in $h_{ij}$ of expansion the expression
     $$\alpha \sqrt {\frac {g}{\tilde g}} \left( \tilde g_{ik}  \tilde
g_{jm} - \frac {1}{2} \tilde g_{ij} \tilde g_{km}\right),$$
     $T_{ij}^{(el)}$ is   the   energy-momentum   tensor  of  averaged
electromagnetic field $F_{ij}$,  $\left(  T_{ij}^{(el)}\right)^{(1)}$
is the sum of all terms of expansion
$\tilde  T_{ij}^{(el)}$
that are first-order in  $h_{ij}$ and  $\omega_{ij}$,
$\left( T_{ij}^{(el)}\right)^{(2)}$   is  the  sum  of  all  terms  of
expansion $\tilde  T_{ij}^{(el)}$
that are second-order in  $h_{ij}$ and  $\omega_{ij}$.

     This expressions has the forms:
    \beq \label {24}
R_{ij}^{(1)} = \nabla_m \Omega^{(1)m}_{ij} - \nabla_j \Omega^{(1)m}_{im},
\eeq
\beq \label {25}
R_{ij}^{(2)} = \nabla_m \Omega^{(2)m}_{ij} - \nabla_j \Omega^{(2)m}_{im}
+ \Omega^{(1)m}_{mn} \Omega^{(1)n}_{ij} -
\Omega^{(1)m}_{jn} \Omega^{(1)n}_{im},
\eeq
где
\beq \label {26}
\Omega^{(1)m}_{ij} = \frac {1}{2}g^{ml}
(-\nabla_l h_{ij} +\nabla_i h_{lj} + \nabla_j h_{li}),
\eeq
\beq \label {27}
\Omega^{(2)}_{ij} = - \frac {1}{2} h^{ml}
(-\nabla_l h_{ij} +\nabla_i h_{lj} + \nabla_j h_{li}) =
-\frac {1}{2} h^m_l \Omega^{(1)l}{ij}.
\eeq

  $$L^{(1)}_{ijkm} = - \frac {1}{2}\left( h_{st} u^s u^t +
    h_{st}g^{st}\right)\left( g_{ik}g_{jm} -
    \frac {1}{2} g_{ij} g_{km} \right) + $$
       \beq \label {28}
       + h_{ik}g_{jm} + g_{ik}h_{jm} - \frac {1}{2}h_{ij}g_{km} -
       \frac {1}{2} g_{ij}h_{km},
    \eeq
       \beq \label {29}
       T_{ij}^{(el)} = \frac {1}{4\pi}\left( - F_{il} F_{jm} g^{lm}  +
\frac {1}{4} g_{ij} F_{lm} F^{lm} \right),
     \eeq
$$\left( T_{ij}^{(el)}\right)^{(1)}  =  \frac  {1}{4\pi}  \Big[   F_{il}
F_{im} h^{lm}  +  \frac  {1}{4}  h_{ij}  F_{lm}  F^{lm} - \frac {1}{2}
g_{ij} F_{lm} F_{kn} g^{lk} h^{mn} - $$
       \beq \label {29a}
     - \omega_{il}  F_{j}^{.l} - \omega_{jl} F_{i}^{.l} + \frac {1}{2}
F^{lm} \omega_{lm} \Big],
     \eeq
$$\left( T_{ij}^{(el)}\right)^{(2)} = \frac {1}{4\pi} \Big[  -  F_{il}
F_{jm} g_{st}  +  \frac  {1}{2}  g_{ij} F_{lk} F_m^{.k} g_{st} + \frac
{1}{4} g_{ij} F_{lm} F_{st} - $$
     $$- \frac  {1}{2}  F_{.t}^{k}  F_{km}  g_{il}  g_{js} \Big] h^{ls}
h^{tm}+$$
     $$+\frac {1}{4\pi}  \Big[  F_{it}  g_{jl}  g_{sm} + F_{jt} g_{il}
g_{sm} - g_{ij} F_{lm} g_{st} +  \frac  {1}{2}  g_{it}  g_{jm}  F_{ls}
\Big] \omega^{ls} h^{tm} + $$
       \beq \label {30}
     +\frac {1}{4\pi}  \left(  -  g_{it}  g_{is} + \frac {1}{4} g_{ij}
g_{st}\right) \omega^{t}_{.l} \omega^{sl}.
     \eeq

     \bigskip

{\bf 2.  Averaging  of microscopic system of Einstein and Maxwell
equations for the relativistic plasma}

We average  (\ref  {22}) and (\ref {23}) over the paths
(Ref. [11]-[14]) and introduce the one-particle distribution function
  \beq \label {31}
  f_a(q,p) = \langle \int d s
\delta^4(q^i -    q^{i}_{a(l)}(s))
\delta^4( p_j - p^{a(l)}_{j}(s))\rangle =  \frac  {1}{n_a}\langle  N_a
\rangle.
\eeq

As a result we have the averaged Einstein equations in the form
     $$R_{ij} +      \Lambda_{ij}      =      \chi(T_{ij}^{(m)}      -
(1/2)T^{(m)}g_{ij})+$$
     \beq \label {33}
+ \chi \left( T_{ij}^{(el)} - \frac {1}{2} T^{(el)} g_{ij} \right) +
+ \chi \left( T_{ij}^{(r)} - \frac {1}{2} T^{(r)} g_{ij} \right).
     \eeq

Here
     \beq \label {34}
 T^{ij}_{(m)}  =  \sum_a  n_a m_a c^2 \int \frac {d^4 p_a}{\sqrt {-g}}
u^i_a u^j_a f_a(q, p_a)-
     \eeq
is the macroscopic energy-momentum tensor of medium,  $T^{ij}_{(el)}$ is  the
energy-momrntum tensor of macroscopic electromagnetic field (see.
(\ref {29}),
     \beq \label {36}
T^{ij}_{(r)} =  \frac {1}{4\pi} \langle - \omega^i_{.l} \omega^{jl} +
\frac {1}{4} g^{ij} \omega_{lm} \omega^{lm} \rangle -
     \eeq
  is the macroscopic energy-momentum tensor of radiation in plasma,
\beq \label {37}
\Lambda_{ij} = \langle R_{ij}^{(2)}\rangle - \sum_a \chi m_a c^2 \int \frac
{d^4 p_a}{\sqrt {-g}} \langle N_a L^{(1)}_{ijkm}u^m_a u^k_a.
\eeq

     When obtained (\ref {33}) we taking into account that
     $$\langle R_{ij}^{(1)}\rangle      =     0, \qquad
\langle     \left( T_{ij}^{(el)}\right)^{(1)} \rangle = 0.$$
     Next we assum that
\beq \label {*}
F_{ik} h^k_j \ll \omega_{ij}\qquad ets.
     \eeq
inside the     correlation     region.      Consequantly      $\langle
\left(T_{ij}^{(el)}\right)^{(2)}\rangle$ is   equal  approximately  to
$T_{ij}^{(r)}$.

Taking into account the  (\ref  {25})  -  (\ref  {28}) we can write
$\Lambda_{ij}$ in the form
\beq \label {38}
\Lambda_{ij} = \nabla_k \varphi^{k}_{ij} + \mu_{ij},
     \eeq
     where
\beq \label {39}
      \varphi^{k}_{ij} = - \frac {1}{2}
\left( \delta^k_n \delta_j^s - \delta_j^k \delta_n^s \right) P^n_{is},
     \eeq
\beq \label {40}
     P^n_{is} = \langle h_l^n \Omega^{(1) l}{is} \rangle,
     \eeq
\beq \label {41}
     \mu_{ij} =  \left( \delta^k_n \delta_j^s - \delta_j^k \delta_n^s
\right) Q^n_{kis} + \lambda_{ij},
     \eeq
\beq \label {42}
     Q^n_{kis} = \langle \Omega^{(1)n}_{kl} \Omega^{(1)l}_{is} \rangle,
     \eeq
$$     \lambda_{ij} = - \sum_{a}\chi m_a c^2\int \frac  {d^4p}{\sqrt
{(-g)}}{\bigl\{} -  \frac{1}{2} u_i u_ju^ku^m - \frac{1}{4}g_{ij}u^ku^m -
\frac{1}{2} u_iu_j g^{km}  + $$
     \beq \label {43}
 + \frac{1}{4} g_{ij}g^{km}  +
u_iu^k\delta^{m}_{j} +u_ju^k\delta^{m}_{i} -
\frac{1}{2}\delta^{k}_{i}\delta^{m}_{j}{\bigr\}} <N_a h_{km}>.
   \eeq

     In (\ref  {43})  we  reject  the  indices  "a"  on  momentums  and
velosities of particles of species "a".

     The macroscopic Maxwell equations, obtained from (\ref  {22})
after averaging, have the form
     \beq \label {44}
    \nabla_k F^{ik} + \nabla_k \varphi^{ik} + \mu^i = -\frac {4\pi}{c}
J^i,
     \eeq
     where
     \beq \label {45}
     \varphi^{ik} = \langle h^i_m \omega^{km}\rangle -
\langle h^k_m \omega^{im}\rangle,
     \eeq
     \beq \label {46}
     \mu^i =  \frac  {1}{2}  \langle \omega^{ik}\nabla_k h\rangle   +
\lambda^i,
     \eeq
     \beq \label {47}
     \lambda^i = - 2\pi \sum_b e_b  \int  \frac  {d^4  p}{\sqrt  {-g}}
u^i_{(b)} \langle N_b h \rangle,
     \eeq
     \beq \label {48}
  J^i = \sum_b e_b c n_b \int \frac {d^4 p}{\sqrt{-g}}u^i_b  f_b -
     \eeq
is the macroscopic current vector.

To simplify still furthe, we only have to calculate
$h_{ij}$ and   $\omega_{ij}$  inside  the  region  determined  by  the
correlations radius and corresponding correlation time. Note, That distant
collisions provide  the  main  contribution  to  caculated macroscopic
quantities. To consider this contribution it is enough to find
$h_{ij}$ and   $\omega_{ij}$   from   the  Einstein-Maxwell  equations
linearised with respect to the metric  $g_{ij}$  and  the  macroscopic
electromagnetic field $F_{ij}$.

We assume the average gravitational field generated by the particles
to be constant within the correlation region.  In  this  case  we  can
interpret $g_{ij}$  within  the correlation region as the Minkowski
metric. We  assume  alsow,   that   the   influence   of   macroscopic
electromagnetical field on the microscopic field in correlation region
is small (see (\ref {*})) .

    As a result we have linearized Einstein and Maxwell equations with
respect the Minkowski metric $g_{ij}$. By employing the gauge
 $\nabla_k  \gamma^{ik}  = 0$, where
$\gamma_{ij}=h_{ij} -  (1/2)h  g_{ij}$,  we get the folloing form of
the linearized Einstein and Maxwell equations
  \beq \label {49}
  \Box \gamma^{ij} = - \sum_b 2\chi m_b c^2 \int d^4 {p'}_b
  \Phi_b(q,{p'}_b) {u'}_b^i {u'}_b^j,
  \eeq
  \beq \label {50}
     \nabla_k \omega^{ik}   =  -  4\pi  \sum_b  e_b  \int  d^4  {p'}_b
\Phi_b(q. {p'}_b) {u'}_b^i,
     \eeq
     Here $\Box = g^{ij} \nabla_i  \nabla_j$,
$\Phi_b = N_b - n_b f_b$.

Thus subsequent calculatios do not have a covariant form, but they are
all done for the purpose of determining the components of the tensors
$\varphi^k_{ij}$, $\mu_{ij}$,     $\varphi_{ij}$,      $\mu_i$      and
$T_{ij}^{(r)}$ at  some  (arbitrary)  point  $(q)$  in   the   locally
Lorentzian frame. In this reference frame the interval $d s^2$ has the form
  \beq \label {51}
     ds^2= d\eta^2 - (dq^1)^2 - (dq^2)^2 -(dq^3)^2.
     \eeq

The final result must be written in covariant form.

     The expressions for $h_{ij}$ and $\Omega^{l}_{ij}$ we get from
the Einstein equations (\ref {49})linearized with respect
to the Minkowski metric
(which we still denote by $g_{ij}$)) were found in ((refer to.[1]):

$$ h_{ij}(\eta,{\bf q}) = \sum_{b} \int d^4 p' \int d^3 q' \int
d^3k \int_{-\infty}^{\eta} d{\eta}'  e^{-i{\bf  k}({\bf  q}-{\bf
 q'})}\times $$
     \beq \label {52}
\times h_{ij}^{(b)} (\eta,{\eta}', p',{\bf k})
\Phi_b ({\eta}',{\bf  q'}, p'),
     \eeq

 $$ \Omega^{i}_{jk} (\eta,{\bf q}) = \sum_{b} \int d^4p' \int d^3
q' \int d^3k \int_{-\infty}^{\eta} d{\eta}'  e^{-i{\bf  k} ({\bf  q}-{\bf
 q'})}\times$$
     \beq \label {53}
\times \Omega^{i(b)}_{jk} (\eta,{\eta}',p',{\bf k})
\Phi_b ({\eta}',{\bf q'}, p').
     \eeq

where (${\bf  q}  =  (q^1,q^2,q^3)$  is the three - dimensional radius
vector in the given reference frame, and ${\bf k} = ( k_1,k_2,k_3)$),
          $$\Phi_b = N_b - n_b f_b $$,
     \beq \label {54}
     h^{(b)}_{ij}(\eta,{\eta}',p',{\bf k}) =
-\frac {i\chi m_b c^2}{(2\pi)^3  k}(u'_{i} u'_{j}
-  \frac {1}{2} g_{ij}) {\Bigl\{} e^{ik( \eta' - \eta)} - e^{-ik(\eta'
- \eta)}{\Bigr\}},
     \eeq
 $$\Omega^{i(b)}_{jk}(\eta,{\eta}',p',{\bf k}) =
\frac {\chi m_b c^2}{2(2\pi)^3  k}{\Bigl\{}{\bigl[}
( u'_{j} u'_{k} -  \frac {1}{2} g_{jk})k^{i}_{+} -
( u'_{j}u'^{i} -  \frac {1}{2} \delta_{j}^{i})k_{k}^{+}-$$
$$- (u'_{k}u'^{i}  -  \frac {1}{2}\delta_{k}^{i})
k_{j}^{+}{\bigr]} e^{ik(\eta' - \eta)} -
{\bigl[}
(u'_{j}u'_{k} -  \frac {1}{2} g_{jk})k^{i}_{-} -
( u'_{j} u'^{i} -  \frac {1}{2} \delta_{j}^{i})k_{k}^{-}-$$
     \beq \label {55}
- ( u'_{k}u'^{i}         -        \frac        {1}{2}
\delta_{k}^{i})k_{j}^{-}{\bigr]} e^{-ik(\eta' - \eta)}{\Bigr\}}.
     \eeq

     In (\ref  {54})  and  (\ref  {55})  the  following  vectors  were
introduced:
$$k^{+}_{i} = ( k, {\bf k}) ,\qquad  k^{-}_{i} = ( -k, {\bf k}), $$

where

$$k = \sqrt {[(k_1)^2 +(k_2)^2 + (k_3)^2]} =
\vert  {\bf k}\vert. $$

Obviously,   $k^{-}_{i}({\bf k})= - k^{+}_{i}(-{\bf k})$.

To obtain the  additional  terms  $\nabla_k  \varphi^{ik}$   and
$\mu^i$ in  macroscopic  Maxwell  equations  we have  to calculate the
$h=h^l_l$ and $\nabla_k h$:

$$ h(\eta,{\bf q}) = \sum_{b} \int d^4 p' \int d^3 q' \int
d^3k \int_{-\infty}^{\eta} d{\eta}'  e^{-i{\bf  k}({\bf  q}-{\bf
 q'})} $$
     \beq \label {56}
h^{(b)} (\eta,{\eta}', p',{\bf k})
\Phi_b ({\eta}',{\bf  q'}, p'),
     \eeq
where
     \beq \label {57}
     h^{(b)}(\eta,{\eta}',p',{\bf k}) =
\frac {i\chi m_b c^2}{(2\pi)^3  k}
{\Bigl\{} e^{ik( \eta' - \eta)} - e^{-ik(\eta'
- \eta)}{\Bigr\}}.
\eeq
$$\nabla_k h(\eta,{\bf q}) = \sum_{b} \int d^4 p' \int d^3 q' \int
d^3k \int_{-\infty}^{\eta} d{\eta}'  e^{-i{\bf  k}({\bf  q}-{\bf
 q'})} $$
     \beq \label {58}
h^{(b)}_{;k} (\eta,{\eta}', p',{\bf k})
\Phi_b ({\eta}',{\bf  q'}, p'),
     \eeq
where
     \beq \label {59}
     h^{(b)}_{;k}(\eta,{\eta}',p',{\bf k}) =
\frac {\chi m_b c^2}{(2\pi)^3  k}
{\Bigl\{} k^{+}_k e^{ik( \eta' - \eta)} - k^{-}_ke^{-ik(\eta'
- \eta)}{\Bigr\}}.
\eeq

Let's write the solution of (\ref {50}) in the form
    $${\omega}_{ik} = {\pa}_i  A_k - {\pa}_k  A_i. $$

     where [11]:
$$ A_i(\eta,{\bf q}) = \sum_{b} \int d^4 p' \int d^3 q' \int
d^3k \int_{-\infty}^{\eta} d{\eta}'  e^{-i{\bf  k}({\bf  q}-{\bf
 q'})} $$
     \beq \label {60}
A_i^{(b)} (\eta,{\eta}', p',{\bf k})
\Phi_b ({\eta}',{\bf  q'}, p').
     \eeq
Here
     \beq \label {61}
     A_i^{(b)}(\eta,{\eta}',p',{\bf k}) =
\frac {i e_b}{(2\pi)^2  k}u'_i
{\Bigl\{} e^{ik( \eta' - \eta)} - e^{-ik(\eta'
- \eta)}{\Bigr\}}.
\eeq

 The $\omega_{ik}$ have the form:
$$ \omega_{ik}(\eta,{\bf q}) = \sum_{b} \int d^4 p' \int d^3 q' \int
d^3k \int_{-\infty}^{\eta} d{\eta}'  e^{-i{\bf  k}({\bf  q}-{\bf
 q'})}\times $$
     \beq \label {62}
\times \omega_{ik}^{(b)} (\eta,{\eta}', p',{\bf k})
\Phi_b ({\eta}',{\bf  q'}, p'),
     \eeq
where
     \beq \label {63}
     \omega_{ik}^{(b)}(\eta,{\eta}',p',{\bf k}) =
\frac { e_b}{(2\pi)^2  k}
{\Bigl\{}(k^{+}_i u'_k - k^{+}_k u'_i) e^{ik( \eta' - \eta)} -
(k^{-}_i u'_k - k^{-}_k u'_i)e^{-ik(\eta'
- \eta)}{\Bigr\}}.
\eeq

  If substitute the  (\ref {52}), (\ref {53}), (\ref {56}),
(\ref {58}),  (\ref {62}) to (\ref {40}),  (\ref  {42}),  (\ref  {43}),
(\ref {45})  ---  (\ref {47}) we get the folloing expressions for
$ P^{n}_{is}$ , $Q^{n}_{kis}$,
$ \lambda_{ij}$, $\langle h^i_m \omega^{km}\rangle$,
$\langle \omega^{ik}\nabla_k h \rangle$, $\lambda_i$:

$$ P^{n}_{is} = \sum_{bc}\int d^4 p'\int d^4p''\int d^3 q'
\int   d^3   q''\int_{-\infty}^{\eta}   d {\eta}'
\int_{-\infty}^{\eta}d{{\eta}''}\int d^3 k' \int d^3 k''\times$$
     \beq \label {64}
\times e^{-i{\bf k'}({\bf q}-{\bf  q'})}
e^{-i{\bf      k''}({\bf      q}-{\bf     q}'')}
 h^{n(b)}_{l}(\eta,{\eta}', p',{\bf k'})\Omega^{l(c)}_{is}
(\eta,{\eta}'',p'',{\bf k''})  n_b  n_c  g_{bc}(x',x''),
\eeq
$$ Q^{n}_{kis} = \sum_{bc}\int d^4 p'\int d^4p''\int d^3 q'
\int   d^3   q''\int_{-\infty}^{\eta}   d {\eta}'
\int_{-\infty}^{\eta}d{{\eta}''}\int d^3 k' \int d^3 k''\times$$
     \beq \label {65}
\times e^{-i{\bf k'}({\bf q}-{\bf  q'})}
e^{-i{\bf      k''}({\bf      q}-{\bf     q}'')}\Omega^{n(b)}_{kl}
(\eta,{\eta}', p',{\bf k'})\Omega^{l(c)}_{is}
(\eta,{\eta}'',p'',{\bf k''})  n_b  n_c  g_{bc}(x',x''),
\eeq
$$ \lambda_{ij} = -\sum_{bc} \chi m_c c^2\int d^4 p'\int d^4p''
\int d^3 q' \int_{-\infty}^{\eta}   d {\eta}'
\int d^3 k' e^{-i{\bf k'}({\bf q}-{\bf  q'})}\times $$
$$\times {\Bigl\{} -\frac{1}{2}{u''}_i  {u''}_j  {u''}^k  {u''}^m  -
\frac{1}{4}g_{ij}{u''}^k {u''}^m   -
\frac{1}{2}{u''}_i{u''}_jg^{km}+\frac{1}{4}g_{ij}g^{km}+{u''}_i {u''}^k
\delta^{m}_{j} + $$
     \beq \label {66}
+ {u''}_j {u''}^k\delta^m_i -\frac{1}{2}\delta^{k}_{i}\delta^{m}_{j}{\Bigr\}}
h^{(b)}_{km}(\eta,{\eta}',p',{\bf k}')n_b  n_c g_{bc}(x';  \eta,  {\bf
q}, p''),
\eeq
$$ \langle h^i_m \omega^{km}\rangle = \sum_{bc}\int d^4 p'\int d^4p''
\int d^3 {\bf q'}\int   d^3   {\bf q''}\int_{-\infty}^{\eta}   d {\eta}'
\int_{-\infty}^{\eta}d{{\eta}''}\int d^3 {\bf k'} \int d^3 {\bf k''}$$
     \beq \label {67}
e^{-i{\bf k'}({\bf q}-{\bf  q'})}
e^{-i{\bf      k''}({\bf      q}-{\bf     q}'')}
 h^{i(b)}_{m}(\eta,{\eta}', p',{\bf k'})\omega_{(c)}^{km}
(\eta,{\eta}'',p'',{\bf k''})  n_b  n_c  g_{bc}(x',x''),
\eeq
$$ \langle \omega^{ik} \nabla_k h \rangle = \sum_{bc}\int d^4 p'\int d^4p''
\int d^3 {\bf q'}\int   d^3   {\bf q''}\int_{-\infty}^{\eta}   d {\eta}'
\int_{-\infty}^{\eta}d{{\eta}''}\int d^3 {\bf k'} \int d^3 {\bf k''}$$
     \beq \label {68}
e^{-i{\bf k'}({\bf q}-{\bf  q'})}
e^{-i{\bf      k''}({\bf      q}-{\bf     q}'')}
 h^{(b)}_{;k}(\eta,{\eta}', p',{\bf k'})\omega_{(c)}^{ik}
(\eta,{\eta}'',p'',{\bf k''})  n_b  n_c  g_{bc}(x',x''),
\eeq
$$ \lambda^i  =  -\sum_{bc} 2\pi e_c\int d^4 p'
\int d^4p''\int d^3 {\bf q'}\int_{-\infty}^{\eta}   d {\eta}'
 \int d^3 {\bf k'}e^{-i{\bf k'}({\bf q}-{\bf  q'})} {u''}^i$$
     \beq \label {69}
h^{(b)}(\eta,{\eta}', p',{\bf k'}) n_b n_c g_{bc}(x';  \eta,  {\bf q},
p'').
\eeq

In this expressions unprimed,  primed  and  double-primed  quantities
refer to  particles  belonging  to  species  "a",  "b" and spesies "c"
respectively.

       In (\ref {64}) - (\ref {69}) we introduced  the  two  -particle
correlation function $g_{ab}(x', x'')$ (see Ref. [1], [2], [11], [12]):

\beq \label {70}
     f_{ab}(x, x')  =  f_a(x)  f_b(x')   +   g_{ab}(x,   x').
     \eeq
     Here $f_a (x)$,  $f_{ab}(x,  x')$,  $f_{abc}(x, x', x'')$ are the
one-particle, two-particle  and  tree-particle  distribution functions
respectively:
     $$\langle \int ds \delta(x -x_a(s)) \rangle = f_a(x),$$
     $$\langle \int ds \delta(x -x_a(s)) \int ds'
\delta(x' -x_b(s')) \rangle = f_{ab}(x, x'),$$
     $$\langle \int ds \delta(x -x_a(s)) \int ds'
\delta(x' -x_b(s')) \int ds'' \delta(x'' -x_c(s''))\rangle =
f_{abc}(x, x', x''),$$

     where

     $$ \delta(x -x_a(s)) = \delta^4(q^i -  q_a^i(s))  \delta^4(p_j  -
p_j^a(s)) .$$

We denote the set of all variable $(\eta,  {\bf q}, p_i)$ by $x$,
the set  $(\eta',  {\bf  q'},  p'_i)$  by  $x'$,  while  the   momenta
$p'_{(b)}$ are denoted by $p'$, and the $p''_{(c)}$ by $p''$.

For the  moments  of  random functions we have the formulas (Ref.[11],
[12]):
\beq \label {73'}
     \langle N_a(x)\rangle = n_a f_a(x),
\eeq
     $$\langle N_a(x)N_b(x')\rangle     =     (n_a     n_b     -    n_a
\delta_{ab})f_{ab}(x, x') + $$,
\beq \label {73}
     + n_a\delta_{ab}f_a(x)\int ds' \delta (x' - x_a(s'/x)),
     \eeq
     $$\langle N_a(x)N_b(x')N_c(x'')\rangle = (n_a n_b n_c - n_a n_b
\delta_{ac} - n_a  n_b  \delta_{bc}  - $$
$$- n_a  n_c \delta_{ab} + 2n_a\delta_{ab}\delta_{bc}) f_{abc}(x,  x',
x'') + $$
$$ + (n_a n_c  - n_a \delta_{ac}) \delta_{ab} f_{ac}(x,  x'')\int ds' \delta
(x' - x_a(s'/x)) +$$
$$+(n_a n_b  - n_a \delta_{ab}) \delta_{ac} f_{ab}(x,  x')\int ds'' \delta
(x'' - x_a(s''/x)) +  $$
$$+(n_a n_b  - n_a \delta_{ab}) \delta_{bc} f_{ab}(x,  x')\int ds'' \delta
(x'' - x_b(s''/x')) +  $$
\beq \label {73a}
     + n_a\delta_{ab}\delta_{bc}f_a(x)\int ds' \delta (x' - x_a(s'/x))
\int ds'' \delta (x'' - x_a(s''/x)).
     \eeq
Here $x_a(s/x)$  stands for the particle path through point $x$ of the
phase space. Bearing in mind that $\Phi_a=N_a -  n_af_a$ and that $f_a$
     is not  a  random function,  we can easily obtain expressions for
the avereges
     $$\langle N_a(x) \Phi_b(x') \rangle, \qquad
\langle N_a(x) \Phi_b(x') \Phi_c(x'')\rangle.$$

In deriving  (\ref {64}) --- (\ref {69}) we assumed that $n_a \gg 1$
and that $x'' = x_b(s''/x')$,  i.e.  point $x''$ is not on the path of
particles of spesies "b" passing through the point $x'$ of  the  phase
spase.

     In work  [2]  two-particles correlation functions $g_{ab}(x',  x'')$
are found for the system gravitationally  interacting  particles.  The
two-particles correlation     functions     the     for    system    of
electromagnetically interacting particles were found by the  author  in
Ref. [13],  when  getting  the  relativistic  kinetic equation for the
plasma (Eq. (18) from Ref. [13]).

     In our case we should find two-particles correlation function
$g_{ab}(x, x')$   caused   by   electromagnetic   and   gravitationall
interactions simultaneously.

     To obtain correlation function inside the correlation  region  we
assume that $h_{ik} \ll 1$, $\omega_{ik} \ll 1$. Therefore
$\Omega ^k_{ij} \simeq  \Omega^{k(1)}_{ij}$,
$\psi^l_{.k} \simeq \omega^l_{.k}$.

After substituting (\ref {53}), (\ref {62}) into
(\ref {16}), multiplying  (\ref  {16})  by  $\Phi_b  (x')$  and
averaging we get
    $$ p^i\frac{\pa   }{\pa   {q^i}}\langle N_a(x) \Phi_b(x')\rangle    +
\Gamma_{j,ik} p^j   p^k\frac{\pa  }{\pa  p_i}\langle N_a(x)\Phi_b(x')\rangle
+ \frac  {e_a}{c}F_{jk}p^k  \frac  {\pa  }{\pa  p_j}  \langle N_a(x)
\Phi_b (x') \rangle = $$
$$=\frac  {\pa}{\pa p_i} \bigg\{\sum_c \int d^4 p''_b
\int d^3  {\bf q''} \int d^3 {\bf k}\int_{-\infty}^{\eta} d\eta''
\exp[-i{\bf k(q-q')}]\times $$
\beq \label {74}
\times \Big[\Omega_{lm}^{j(c)}(\eta, \eta'',  p''_b,  {\bf k})
p^l p^m  \Delta_{ji} - \frac {e_a}{c} \omega^{l(c)}_{.k}(\eta,  \eta',
p'', {\bf                k})p^k                \Delta_{lj}\Big]\langle
N_a(x)\Phi_b(x')\Phi_c(x'')\rangle \bigg\}.
     \eeq

Next we assume that
     $$f_{abc}(x, x', x'') \simeq f_a(x) f_b(x'), f_c(x'')$$.

In view of Eqs.(\ref {73}),  (\ref {73a}) the two partial  correlation
function  $g_{ab}(x, x')$ obeys the equation
$$ p^i\frac{\pa }{\pa {q^i}}g_{ab}(x,x')  +  \Gamma_{i,jk}p^i  p^k
\frac {\pa}{\pa  p_j} g_{ab}(x,  x') + \frac {e_a}{c} F_{jk} p^k \frac
{\pa}{\pa p_j} g_{ab} (x, x') = $$
$$ = \frac  {\pa}{\pa p_i} \bigg\{\int d^4 p''_b \int d^3  {\bf q''}
\int d^3 {\bf k}\int_{-\infty}^{\eta} d\eta''\exp[-i{\bf k(q-q'')}]\times $$
$$\times \Big[ \Omega_{lm}^{j(b)}(\eta, \eta'',  p'',  {\bf k})
p^l p^m  \Delta_{ji} - \frac {e_a}{c} \omega^{(c)}_{jk}(\eta,  \eta'',
p'', {\bf k}) p^k \Big] \times $$
\beq \label {75}
\times f_a(x)  f_b(x')  \int  ds''  \delta  (x''  -
x_b(s''/x'))\bigg\}.
     \eeq

In this equation for $g_{ab}(x, x')$ we should put  $\Gamma_{i,jk}=0$,
$F_{jk}=0$, since we assume that within  the  correlation  region  the
metric $g_{ij}$  are  constant  and  that the influence of macroscopic
electromagnetic field is small.

     So we have the first order linear equation (\ref {75}),  whose
right-hand side  contains the sum of two terms.  The first term caused
by gravitationall   interactions,   the   second   one    caused    by
electromagnetic interaction.

     Consequently, we  can  write the solution of Eqs.  (\ref {75}) in
the form

\beq \label {76}
     g_{ab}(x, x') = g_{ab}^{(gr)} (x, x') + g_{ab}^{(el)}(x, x').
     \eeq
     Here $g_{ab}^{(gr)} (x, x')$ caused by gravitational interaction,
and  $g_{ab}^{(el)} (x, x')$  caused by electromagnetic interaction.

     The equation for $g_{ab}^{(gr)} (x,  x')$ coinside with  one  in
Ref. [1].  In [1] we got the $g_{ab}^{(gr)} (x,  x')$ in the form (see
Eq. (46) in Ref. [1]):
     $$g_{ab}^{(gr)}(x,x')=
\int d^3 k \int_{-\infty}^{\eta}\frac{d{\tau}}{p^0}
    {\Bigl[}\frac{\pa}{\pa p_i}(p^lp^m\Delta_{ij}f_{a}(x)){\Bigr]}_{\tau}
\int_{-\infty}^{\tau}\frac{d{\tau}'}{u'^0}f_b(x')\times $$
$$ \times {\Omega}^{j(b)}_{lm}(\tau, \tau', p', {\bf k})
exp[-i{\bf k}({\bf q}-{\bf q'})
+   \frac{i}{c}({\bf k}{\bf v})(\eta-\tau) +
     \frac{i}{c}({\bf k}{\bf v'})({\tau}'-{\eta}')] + $$
     $$ + \int d^3 k \int_{-\infty}^{\eta'}\frac{d{\tau'}}{{p'}^0}
    {\Bigl[}\frac{\pa}{\pa {p'}_i}({p'}^l{p'}^m{\Delta'}_{ij}
f_{b}(x')){\Bigr]}_{\tau'}
\int_{-\infty}^{\tau}\frac{d{\tau}}{u^0}f_a(x)\times $$
\beq \label {77}
\times {\Omega}^{j(a)}_{lm}(\tau', \tau, p, {\bf k})
exp[-i{\bf k}({\bf q}-{\bf q'})
+   \frac{i}{c}({\bf k}{\bf v'})(\eta'-\tau') +
     \frac{i}{c}({\bf k}{\bf v})(\tau-\eta)],
\eeq

where ${\bf v} = c{\bf u}_a/u^0_a$, ${\bf v'} = c{\bf u'}_b/{u'}^0_b$,
$ {\bf u}_a = (u^1,  u^2,  u^3)$,  $ {\bf  u'}_b  =  ({u'}^1,  {u'}^2,
{u'}^3)$.
Here the subscript $\tau$ indicates that  after  calculating
the derivatives with respect $p$ we must replace the arguments
$\eta$,  and ${\bf q}$ by  $\tau$ and $ {\bf q} + \frac {\bf v}{c} (\tau -
\eta)$, respectively.
     The subscript $\tau'$ indicates that  after  calculating
     the derivatives with respect $p'$ we must replace the arguments
     $\eta'$,  and ${\bf q'}$ by  $\tau'$ and $ {\bf q'} +
\frac {\bf v'}{c} (\tau' - \eta')$, respectively.

The equation for $g_{ab}^{(el)}(x,  x')$ coinside with one in Ref.[13].
In Ref. [13] we got the solution for $g_{ab}^{(el)}(x,  x')$
(See (18) from Ref. [13]).

With the preceding notation the result (18) from Ref. [13] takes the form:

    $$g_{ab}^{(el)}(x, x')= -\frac {e_b e_c}{2\pi^2 c}\int \frac {d^3 k}{k}
\int_{-\infty}^{\eta'}d\tau'\frac {u'^k}{u'^0}f_c(x'')
\frac{\pa f_b(x')}{\pa p'_i}\int_{-\infty}^{\tau'}d\tau''\times$$
$$\times {\biggl\{}\frac {u''_k}{u''^0} \frac {\pa}{q'^{i}_{\tau'}}
{\lbrack}exp(-i{\bf k}({\bf      q'}-{\bf      q''})sin(k(\tau'      -
\tau'')){\rbrack} - $$
$$ - \frac {u''_i}{u''^0} \frac {\pa}{q'^{k}_{\tau'}}
{\lbrack}exp(-i{\bf k}({\bf      q'}-{\bf      q''})sin(k(\tau'       -
\tau'')){\rbrack}{\biggr\}}\times $$
$$ \times exp[ \frac{i}{c}({\bf k}{\bf v'})(\eta'-\tau') +
     \frac{i}{c}({\bf k}{\bf v''})({\tau}''-{\eta}'')]- $$
    $$ -\frac {e_b e_c}{2\pi^2 c}\int \frac {d^3 k}{k}
\int_{-\infty}^{\eta''}d\tau''\frac {u''^k}{u''^0}f_b(x')
\frac{\pa f_c(x'')}{\pa p''_i}\int_{-\infty}^{\tau''}d\tau'\times$$
$$\times {\biggl\{}\frac {u'_k}{u'^0} \frac {\pa}{q''^{i}_{\tau''}}
{\lbrack}exp(-i{\bf k}({\bf     q''}-{\bf      q'})sin(k(\tau''      -
\tau')){\rbrack} - $$
$$ - \frac {u'_i}{u'^0} \frac {\pa}{q''^{k}_{\tau''}}
{\lbrack}exp(-i{\bf k}({\bf      q''}-{\bf      q'})sin(k(\tau''       -
\tau')){\rbrack}{\biggr\}}\times $$
     \beq \label {78}
     \times exp[ \frac{i}{c}({\bf k}{\bf v''})(\eta''-\tau'') +
     \frac{i}{c}({\bf k}{\bf v'})({\tau}'-{\eta}')].
     \eeq

Here $q^{i}_{\tau}  =  (\tau,  q^{  \alpha})$.

After performing a differentiation with respect to $q^{i}_{\tau}$ in
(\ref  {26}) we get a following
expression for correlation function:

    $$g_{ab}^{(el)}(x, x')= -\frac {e_b e_c}{4\pi^2 c}\int \frac {d^3 k}{k}
\int_{-\infty}^{\eta'}\frac {d\tau'}{u'^0}f_c(x'')
\frac{\pa f_b(x')}{\pa p'_i}\int_{-\infty}^{\tau'}\frac {d\tau''}{u''^0}
e^{ -i{\bf k}({\bf      q'}-{\bf      q''})}
{\lbrack}(u'u'')\delta^{k}_{i} -  $$
$$ -u''_i u'^k{\rbrack}
\left( k_{k}^{+}e^{ik(\tau'' - \tau')} -
k_{k}^{-}e^{-ik(\tau'' - \tau')}\right)
exp[ \frac{i}{c}({\bf k}{\bf v'})(\eta'-\tau') +
     \frac{i}{c}({\bf k}{\bf v''})({\tau}''-{\eta}'')]- $$
    $$ -\frac {e_b e_c}{4\pi^2 c}\int \frac {d^3 k}{k}
\int_{-\infty}^{\eta''}\frac {d\tau''}{u''^0}f_b(x')
\frac{\pa f_c(x'')}{\pa p''_i}\int_{-\infty}^{\tau''}\frac {d\tau'}{u'^0}
e^{-i{\bf k}({\bf      q''}-{\bf      q'})}
{\lbrack} (u'u'')\delta^{k}_{i} - $$
     \beq \label {79}
-u'_i u''^k{\rbrack}\left( k_{k}^{+}e^{ik(\tau' - \tau'')} -
k_{k}^{-}e^{-ik(\tau' - \tau'')}\right)
exp[ \frac{i}{c}({\bf k}{\bf v''})(\eta''-\tau'') +
     \frac{i}{c}({\bf k}{\bf v'})({\tau}'-{\eta}')].
     \eeq

    It is evident, that  the  electromagnetic  interactios  in  plasma are
dominating. Consequently
      $$g_{ab}^{(gr)} (x, x') \ll g_{ab}^{(el)}(x, x').$$
That is  why  one  can  put  the  $g_{ab}^{(el)}(x,  x')$  instead  of
$g_{ab}(x, x')$ in  (\ref  {64})  - (\ref {69}).

     If we now substitude (\ref {79}) into (\ref {64}) - (\ref {69}) and
integrate with respect  ${\bf  q'}$,  ${\bf q''}$,  ${\bf k'}$,
${\bf k''}$ we get the following expressions:

$$ P^{n}_{is}  =  \sum_{bc}\left(  -\frac  {e_b  e_c  n_b  n_c (2\pi)^6}
{4\pi^2c}\right)\int d^4 p'\int d^4p''\int_{-\infty}^{\eta}   d {\eta}'
\int_{-\infty}^{\eta}d{\eta}''\int \frac {d^3k}{k}\times$$
$$\times {\biggl\{}\int_{-\infty}^{{\eta}'}\frac{d{\tau}'}{u'^0}
\int_{-\infty}^{\tau'}\frac{d{\tau}''}{u''^0}
f_c(x'')\frac{\pa f_b(x')}{\pa p'_j}
{\lbrack}(u'u'')\delta^{k}_{j} - u''_j u'^k{\rbrack}\times$$
$$\times \left( - k_{k}^{-}e^{ik(\tau'' - \tau')} +
k_{k}^{+}e^{-ik(\tau'' - \tau')}\right)exp[ \frac{i}{c}({\bf k}{\bf v''})
(\eta''-\tau'') + \frac{i}{c}({\bf k}{\bf v'})({\tau}'-{\eta}')]+ $$
$$ + \int_{-\infty}^{{\eta}''}\frac{d{\tau}''}{u''^0}
\int_{-\infty}^{\tau''}\frac {d\tau'}{u'^0}
f_b(x') \frac {\pa f_c(x'')}{\pa p''_j}
{\lbrack}(u'u'')\delta^{k}_{j} - u'_j u''^k{\rbrack}\times$$
$$\times \left(  k_{k}^{+}e^{-ik(\tau'' - \tau')} -
k_{k}^{-}e^{ik(\tau'' - \tau')}\right)
exp[ \frac{i}{c}({\bf k}{\bf v''})(\eta''-\tau'') +
     \frac{i}{c}({\bf k}{\bf v'})({\tau}'-{\eta}')]{\biggr\}}\times$$
   \beq \label {80}
\times h^{n(b)}_{l}(\eta,\eta', p',-{\bf k})
\Omega^{l(c)}_{is}(\eta,{\eta}'',p'',{\bf k}),
\eeq

$$ Q^{n}_{kis}  =  \sum_{bc}\left(  -\frac  {e_b  e_c  n_b  n_c (2\pi)^6}
{4\pi^2c}\right)\int d^4 p'\int d^4p''\int_{-\infty}^{\eta}   d {\eta}'
\int_{-\infty}^{\eta}d{\eta}''\int \frac {d^3k}{k}\times$$
$$\times {\biggr\{}\int_{-\infty}^{{\eta}'}\frac{d{\tau}'}{u'^0}
\int_{-\infty}^{\tau'}\frac{d{\tau}''}{u''^0}
f_c(x'')\frac{\pa f_b(x')}{\pa p'_j}
{\lbrack}(u'u'')\delta^{k}_{j} - u''_j u'^k{\rbrack}\times$$
$$\times \left( - k_{k}^{-}e^{ik(\tau'' - \tau')} +
k_{k}^{+}e^{-ik(\tau'' - \tau')}\right)exp[ \frac{i}{c}({\bf k}{\bf v''})
(\eta''-\tau'') + \frac{i}{c}({\bf k}{\bf v'})({\tau}'-{\eta}')]+ $$
$$ + \int_{-\infty}^{{\eta}''}\frac{d{\tau}''}{u''^0}
\int_{-\infty}^{\tau''}\frac {d\tau'}{u'^0}
f_b(x') \frac {\pa f_c(x'')}{\pa p''_j}
{\lbrack}(u'u'')\delta^{k}_{j} - u'_j u''^k{\rbrack}\times$$
$$\times \left(  k_{k}^{+}e^{-ik(\tau'' - \tau')} -
k_{k}^{-}e^{ik(\tau'' - \tau')}\right)
exp[ \frac{i}{c}({\bf k}{\bf v''})(\eta''-\tau'') +
     \frac{i}{c}({\bf k}{\bf v'})({\tau}'-{\eta}')]{\biggr\}}\times$$
   \beq \label {81}
\times \Omega^{n(b)}_{kl}(\eta,\eta', p',-{\bf k})
\Omega^{l(c)}_{is}(\eta,{\eta}'',p'',{\bf k}),
\eeq

$$ \lambda_{ij}   =   \sum_{bc}\chi (2\pi)  e_b  e_c  n_b n_c  m_c c
\int d^4 p'\int d^4p''
{\lbrack} -\frac{1}{2} u''_{i} u''_{j} u''^{k} u''^{m}  -
\frac{1}{4}g_{ij}u''^k u''^m   - $$
$$ -\frac{1}{2}u''_i u''_j g^{km} + \frac{1}{4}g_{ij}g^{km}
 + u''_i u''^k \delta^{m}_{j} + u''_j u''^k \delta^{m}_{i} -
\frac{1}{2}\delta^{k}_{i}\delta^{m}_{j}{\rbrack}\times$$
$$\times \int \frac {d^3 k}{k} \int_{-\infty}^{\eta}   d \eta'
 h^{(b)}_{km}(\eta,\eta',p',-{\bf k})
 {\bigg\{}\int_{-\infty}^{\eta'}\frac{d\tau'}{u'^0}
 \int_{-\infty}^{\tau'}\frac{d\tau''}{u''^0}f_c(x'')\frac {\pa f_b(x')}
{\pa p'_r}\times $$
$$\times \left( (u'u'')\delta^{l}_{r} - u''_r u'^l\right)
 \left(  k_{l}^{+}e^{-ik(\tau'' - \tau')} -
k_{l}^{-}e^{ik(\tau'' - \tau')}\right)\times $$
$$\times exp[\frac{i}{c}({\bf k}{\bf v''})(\eta -\tau'') +
     \frac{i}{c}({\bf k}{\bf v'})(\tau'-\eta')] + $$
$$ + \int_{-\infty}^{\eta}\frac{d\tau''}{u''^0}
 \int_{-\infty}^{\tau''}\frac{d\tau'}{u'^0}f_b(x')\frac {\pa f_c(x'')}
{\pa p''_r}      \left(       (u'u'')\delta^{l}_{r}       -       u'_r
u''^l\right)\times$$
     \beq \label {82}
\times \left(  k_{l}^{+}e^{-ik(\tau'' - \tau')} -
k_{l}^{-}e^{ik(\tau'' - \tau')}\right)
exp[\frac{i}{c}({\bf k}{\bf v''})(\eta -\tau'') +
     \frac{i}{c}({\bf k}{\bf v'})(\tau'-\eta')]\bigg\}.
     \eeq

$$ \langle h^i_m \omega^{km}\rangle  =
\sum_{bc} \left(  -\frac  {e_b  e_c  n_b  n_c (2\pi)^6}
{4\pi^2c}\right)\left( \frac   {-i\chi   m_b  c^2}{(2\pi)^3  k}\right)
\left( \frac {e_c}{4\pi^2k}\right)\int d^4 p'\int d^4p''\times $$
$$\times \int_{-\infty}^{\eta}   d {\eta}'
\int_{-\infty}^{\eta}d{\eta}''\int \frac {d^3k}{k}
{\biggl\{}\int_{-\infty}^{{\eta}'}\frac{d{\tau}'}{u'^0}
\int_{-\infty}^{\tau'}\frac{d{\tau}''}{u''^0}
f_c(x'')\frac{\pa f_b(x')}{\pa p'_l}
{\lbrack}(u'u'')\delta^{s}_{l} - u''_l u'^s{\rbrack}\times$$
$$\times \left( - k_{s}^{-}e^{ik(\tau'' - \tau')} +
k_{s}^{+}e^{-ik(\tau'' - \tau')}\right)exp[ \frac{i}{c}({\bf k}{\bf v''})
(\eta''-\tau'') + \frac{i}{c}({\bf k}{\bf v'})({\tau}'-{\eta}')]+ $$
$$ + \int_{-\infty}^{{\eta}''}\frac{d{\tau}''}{u''^0}
\int_{-\infty}^{\tau''}\frac {d\tau'}{u'^0}
f_b(x') \frac {\pa f_c(x'')}{\pa p''_l}
{\lbrack}(u'u'')\delta^{s}_{l} - u'_l u''^s{\rbrack}\times$$
$$\times \left(  k_{s}^{+}e^{-ik(\tau'' - \tau')} -
k_{s}^{-}e^{ik(\tau'' - \tau')}\right)
exp[ \frac{i}{c}({\bf k}{\bf v''})(\eta''-\tau'') +
     \frac{i}{c}({\bf k}{\bf v'})({\tau}'-{\eta}')]{\biggr\}}\times$$
$$\times (u'^i u'_m - \frac {1}{2}\delta^i_m)\left( e^{ik(\eta'-\eta)} -
e^{-ik(\eta'-\eta)}\right) \big[{u''}^m
(\left( k_{+}^k e^{ik(\eta''-\eta)}
- k_{-}^k e^{-ik(\eta''-\eta)}\right) - $$
   \beq \label {83}
- {u''}^k(\left( k_{+}^m e^{ik(\eta''-\eta)}
- k_{-}^m e^{-ik(\eta''-\eta)}\right)\big],
\eeq

$$ \langle  \omega^{ik}\nabla_k h \rangle =
\sum_{bc} \left(  -\frac  {e_b  e_c  n_b  n_c (2\pi)^6}
{4\pi^2c}\right)\left( \frac   {\chi   m_b  c^2}{(2\pi)^3  k}\right)
\left( \frac {e_c}{4\pi^2k}\right)\int d^4 p'\int d^4p''\times $$
$$\times \int_{-\infty}^{\eta}   d {\eta}'
\int_{-\infty}^{\eta}d{\eta}''\int \frac {d^3k}{k}
{\biggl\{}\int_{-\infty}^{{\eta}'}\frac{d{\tau}'}{u'^0}
\int_{-\infty}^{\tau'}\frac{d{\tau}''}{u''^0}
f_c(x'')\frac{\pa f_b(x')}{\pa p'_l}
{\lbrack}(u'u'')\delta^{s}_{l} - u''_l u'^s{\rbrack}\times$$
$$\times \left( - k_{s}^{-}e^{ik(\tau'' - \tau')} +
k_{s}^{+}e^{-ik(\tau'' - \tau')}\right)exp[ \frac{i}{c}({\bf k}{\bf v''})
(\eta''-\tau'') + \frac{i}{c}({\bf k}{\bf v'})({\tau}'-{\eta}')]+ $$
$$ + \int_{-\infty}^{{\eta}''}\frac{d{\tau}''}{u''^0}
\int_{-\infty}^{\tau''}\frac {d\tau'}{u'^0}
f_b(x') \frac {\pa f_c(x'')}{\pa p''_l}
{\lbrack}(u'u'')\delta^{s}_{l} - u'_l u''^s{\rbrack}\times$$
$$\times \left(  k_{s}^{+}e^{-ik(\tau'' - \tau')} -
k_{s}^{-}e^{ik(\tau'' - \tau')}\right)
exp[ \frac{i}{c}({\bf k}{\bf v''})(\eta''-\tau'') +
     \frac{i}{c}({\bf k}{\bf v'})({\tau}'-{\eta}')]{\biggr\}}\times$$
$$\times \left(-k_k^{-} e^{ik(\eta'-\eta)} -
k^{+}_k e^{-ik(\eta'-\eta)}\right) \big[{u''}^k
(\left( k_{+}^i e^{ik(\eta''-\eta)}
- k_{-}^i e^{-ik(\eta''-\eta)}\right) - $$
   \beq \label {84}
- {u''}^i(\left( k_{+}^k e^{ik(\eta''-\eta)}
- k_{-}^k e^{-ik(\eta''-\eta)}\right)\big],
\eeq

$$ \lambda^{i}   =   \sum_{bc}\frac   {i\chi e_b  e^2_c m_b c n_b
n_c}{2\pi}\int d^4 p'\int d^4p''\int_{-\infty}^{\eta}   d \eta'
\int \frac {d^3 k}{k^2} \times$$
$$\times {u''}^i \left( e^{ik(\eta'-\eta)} - e^{-ik(\eta'-\eta)}\right)
{\biggl\{}\int_{-\infty}^{\eta'}\frac{d\tau'}{u'^0}
 \int_{-\infty}^{\tau'}\frac{d\tau''}{{u''}^0}f_c(x'')\frac {\pa f_b(x')}
{\pa p'_l} \Big( (u'u'') \delta^{s}_{l} - $$
$$-{u''}_l u'^s \Big) \left(  k_{s}^{+} e^{-ik(\tau'' - \tau')} -
k_{l}^{-}e^{ik(\tau'' - \tau')}\right)
exp[\frac{i}{c}({\bf k}{\bf v''})(\eta -\tau'') +
     \frac{i}{c}({\bf k}{\bf v'})(\tau'-\eta')] + $$
$$+\int_{-\infty}^{\eta}\frac{d\tau''}{u''^0}
 \int_{-\infty}^{\tau''}\frac{d\tau'}{u'^0}f_b(x')\frac {\pa f_c(x'')}
{\pa p''_l} \left( (u'u'')\delta^{s}_{l} - u'_l u''^s\right)
\Big( k_{s}^{+}e^{-ik(\tau'' - \tau')} - $$
   \beq \label {85}
 - k_{s}^{-}e^{ik(\tau'' - \tau')}\Big)
exp[\frac{i}{c}({\bf k}{\bf v''})(\eta -\tau'') +
     \frac{i}{c}({\bf k}{\bf v'})(\tau'-\eta')]{\biggr\}}.
     \eeq

     To simplify (\ref {80}) --- (\ref {85}) still further, we proceed
as follows.  We assume that the distribution function  changes
little inside the correlation region,  so that in calculating the
integrals in (\ref {80}) --- (\ref {85}) we can  ignore,  in   first
approximation, the  temporal coordinate dependence  on $f$.  We
substitute the explicit  expressions  for  $h_{ij}^{(b)}$  and
$\Omega_{kj}^{i(b)}$ (Eqs. (\ref {54}) and (\ref {55})) into (\ref {80})
- (\ref {82}) and evaluate  the  integrals  with  respect  to  $\tau',
\tau'', \eta',   \eta''$  and  ${\bf  k}$.  Then  the  expression  for
$P^{n}_{is}$ becomes

$$ P_{is}^{n}  =
\sum_{bc}\frac{\chi^2 e_b e_c m_{b} m_{c} n_b n_c c^3}{8(\pi)^2}
\int d^4p' \int d^4p''{\bigl(} u'^n   u'_l    -
\frac{1}{2}\delta^{n}_{l}{\bigr)}
{\Bigl[} (u''_i u''_s - \frac{1}{2} g_{is})g^{lf} - $$
$$ - (u''^l u''_i - \frac{1}{2}\delta^{l}_{i})\delta^{f}_{s} -
(u''^l u''_s - \frac{1}{2}\delta^{l}_{s})\delta^{f}_{i}{\Bigr]}
{\biggl\{}{\Bigl[} (u'u'')\delta^{m}_{j} - u''_j u'^m {\Bigr]}
f_c(x'')\frac{\pa f_b(x')}{\pa p'_j} K_{fm}^{(1)}(u',u'') + $$
     \beq \label {85a}
+ {\Bigl[} (u'u'')\delta^{m}_{j}     -     u'_j     u''^m    {\Bigr]}
f_b(x')\frac{\pa f_c(x'')}{\pa p''_j} K_{fm}^{(2)}(u',u''){\biggr\}}.
\eeq

     Here we have introduced the notation $K_{fm}^{(1)}(u',u'')$ and
$K_{fm}^{(2)}(u',u'')$ for   tensors   that   in   locally  Lorentzian
reference frame,  in which  $g_{ij}  =  \eta_{ij}$  is  the  Minkowski
tensor, have the following form:

$$ K_{fm}^{(1)}(u',u'') = \frac{i}{u'^0 u''^0}
\int\frac{d^3 k}{k^3}\int_{-\infty}^{\eta}   d\eta' \int_{-\infty}^{\eta}
d\eta''\int_{-\infty}^{\eta'}    d\tau' \int_{-\infty}^{\tau'}
d\tau'' {\bigl(} e^{ik(\eta'-\eta)} - $$
$$ - e^{-ik(\eta'-\eta)}{\bigr)}{\bigl(} k^{+}_{f}e^{ik(\eta''-\eta)}            -           k^{-}_{f}
e^{-ik(\eta''-\eta)}{\bigr)}{\bigl(} k^{+}_{m}e^{-ik(\tau''-\tau')}            -           k^{-}_{m}
e^{ik(\tau''-\tau')}{\bigr)}\times$$
$$\times exp[\frac{i}{c}({\bf kv''})(\eta''-\tau'') +
     \frac{i}{c}({\bf kv'})({\tau}'-{\eta}')], $$

$$ K_{fm}^{(2)}(u',u'')  =   \frac{i}{u'^0 u''^0}\int\frac{d^3k}{k^3}
\int_{-\infty}^{\eta}   d\eta' \int_{-\infty}^{\eta}
d\eta''\int_{-\infty}^{\eta''}   d\tau''\int_{-\infty}^{\tau''}
d\tau' {\bigl(} e^{ik(\eta'-\eta)} - $$
$$ - e^{-ik(\eta'-\eta)}{\bigr)}{\bigl(} k^{+}_{f}e^{ik(\eta''-\eta)}           -            k^{-}_{f}
e^{-ik(\eta''-\eta)}{\bigr)}{\bigl(} k^{+}_{m}e^{-ik(\tau''-\tau')}           -            k^{-}_{m}
e^{ik(\tau''-\tau')}{\bigr)}\times$$
$$\times exp[\frac{i}{c}({\bf kv''})(\eta''-\tau'') +
     \frac{i}{c}({\bf kv'})({\tau}'-{\eta}')], $$

     After  carrying out the integrals with respect  $\tau'$, $\tau''$,
 $\eta'$, and $\eta''$ these expressions take the form:
 $$K_{fm}^{(1)}(u',u'') =    \frac{2\pi c^5}{u'^0 u''^0}
\int\frac{d^3 k}{k^2} \delta   {\bf   (kv''-    kv')}
{\biggl\{} \frac{k_{f}^{+}k_{m}^{+}}{(kc- {\bf kv''})(kc+ {\bf kv''})^3} + $$
     \beq \label {85b}
 + \frac{k_{f}^{+}k_{m}^{-} + k_{f}^{-}k_{m}^{+} }
{(kc- {\bf kv''})^2(kc+ {\bf kv''})^2} +
\frac{k_{f}^{-}k_{m}^{-}}{(kc- {\bf    kv''})^3(kc+    {\bf    kv''})}
{\biggr\}} = K_{fm}(u',u''),
    \eeq

     \beq \label {85c}
   K_{fm}^{(2)}(u',u'') = - K_{fm}^{(1)}(u',u'') = - K_{fm}(u',u'')
\eeq

     The above equalities hold only in a locally Lorentzian reference frame.
To obtain covariant expressions for the tensors
$K^{(1)}_{fm}(u',u'')$ and   $K^{(2)}_{fm}(u',u'')  $,  we  take  into
account the following fact. The quantities
$K^{(1)}_{fm}(u',u'')$ and $K^{(2)}_{fm}(u',u'') $ appeared in
(\ref {85a}) after the  correlation function $g_{ab}(x', x'')$
was substituted  to (\ref {64}) and result was integrated with respect
to $ {\bf   q',q'',k'} $ and ${\bf k''}$. But the expression
(\ref {79})  for   two - particle correlation function is a sum of
two terms,  which differ  in  that  primed  quantities  referring  to
particles of  species $"a"$ are replaced by double - primed quantities
referring to particles of species $"b"$, and vice versa.
It is after these terms were integrated with respect
$  {\bf   q',q'',k'}$ and ${\bf k''}$   that
$K^{(1)}_{fm}(u',u'')$ and $K^{(2)}_{fm}(u',u'')$  appeared  in  (\ref
{85a}).
Obviously, the both must be calculated in the  same  reference  frame,
for which  it  is  convenient to take the center - of - mass reference
frame, in which
$${\bf v'} = {\bf v},\qquad {\bf v''} = -{\bf v},\qquad u'^0 = u''^0 =
1/\sqrt {1- v^2/c^2} =u^0 $$.
In this reference frame

     \beq \label {85d}
K_{00} =  K_{0  \alpha}=0,\qquad  K_{\alpha  \beta}  =
\frac  {2 \pi^2 c}{v u_{0}^{2} k_{min}^{2}}\left( \delta_{\alpha   \beta} -
\frac   {v_{\alpha} v_{\beta}}{v^2}\right)
     \eeq
     Here $ v = \sqrt {v_{1}^{2} + v_{2}^{2} + v_{3}^{2}} $ , where $
v_{\alpha} =  v^{\alpha} = u^{\alpha}/u^0 $ are spatial components of
the vector ${\bf v}$.

     A covariant generalization of (\ref {85d}) has the form

$$ K_{ij}(u', u'')    =    \frac    {4    \pi^2}{k_{min}^{2}[(u'u'')^2 -
1]^{3/2}}{\bigl\{} - {\lbrack} (u' u'')^2 -1 {\rbrack} g_{ij} - $$
     \beq \label {85e}
- u'_{i} u'_{j} - u''_{i} u''_{j} + (u' u'')(u'_{i} u''_{j} +
u''_{i}  u'_{j}){\bigr\}}
     \eeq

     The expressions for  $K^{(1)}_{fm}(u',u'')$ and
$K^{(2)}_{fm}(u',u'')$ diverge as
 $ k \rightarrow 0 $, i.e., for large impact parameters. The reason is
that we integrate over an infinite region,  while actually  we  should
integrate only over the correlation region,  where the metric is
assumed to vary only weakly.  This difficulty is resolved,  as well as
in the case of kinetic equation deriving,  by introducing a cutoff procedure
in the divergent integral
     $$ \int_{0}^{\infty} \frac {dk}{k^3} $$.

     We set the lower integration limit to
$ k_{min} = 1/r_{max}$,  rather than zero, where $r_{max}$ is the size
of the correlation region (the correlation radius). Then the above
integral assumes the value $1/2k_{min}^{2} = (1/2)r^{2}_{max}$.

     As the experience of deriving the relativistic  kinetic  equation
     (refer to. [13],  [14],  [16], [20]) shows, more thorough investigations
suggest that the integrals become convergent as $r \rightarrow \infty $,
with the contribution from the region where $ r > r_{max} $ being
infinitesimal. In Ref. [14],[16] there are estimates for
$r_{max}$ in the case
where the  average metric $g_{ij}$ is the metric of isotropic
cosmological model and in the case of gravitational interaction of particles.

     In the case of  electromagnetical interaction of particles the
parameter $k_{min}$  is equal to $\frac {1}{r_{D}}$, where $ r_{D}$
a radius of Debit, since electromagnetic interactions in the
plasma are shielded under $r>r_{D}$.

     The tensor (\ref {85e}) possesses the following properties:
\beq \label {85f}
K_{ij}(u', u'') = K_{ij}(u'',  u') ;  K_{ij} u'^{i} = K_{ij} u''^{i} =
0; K_{ij} = K_{ji}.
      \eeq
Because of this the expression for $P^{n}_{is}$ simplifies
considerably. The macroscopic Einstein equations incorporate not
$P^{n}_{is}$, but the tensor $ \varphi^{k}_{ij} =
- (1/2)(\delta^{k}_{n}\delta^{s}_{j} -
\delta^{k}_{j}\delta^{s}_{n})P^{n}_{is}$.
The expression for this tensor can be written as follows:

$$\varphi^{k}_{ij} =
- (1/2)(\delta^{k}_{n}\delta^{s}_{j} -
\delta^{k}_{j}\delta^{s}_{n})P^{n}_{is} =
 \sum_{bc} \frac {\chi^{2} e_b e_c m_{b} m_{c} n_b
n_c c^3 }{16( \pi)^2}\int \frac {d^4 p'}{\sqrt {(-g)}}\int
\frac {d^4 p''}{\sqrt {(-g)}}{\biggl[}
\frac{1}{2} g^{fk} u''_{i} u''_{j} + $$
     \beq \label {86}
 +u'^{k}  (u'  u'')(\delta^{f}_{j}  u''_{i}  +  \delta^{f}_{i}
u''_{j}){\biggr]} (u' u'')  K_{fr}(u',u'')
\left( f_c(x'') \frac {\pa f_b(x')}{\pa p'_{r}} -
f_b(x') \frac {\pa f_c(x'')}{\pa p''_{r}}\right)
     \eeq

     Note that
\beq \label {85h}
g^{ij}\varphi^{k}_{ij} = 0,\qquad \varphi^{i}_{ij} =0,
\qquad \varphi^{k}_{ij} = \varphi^{k}_{ji}.
     \eeq

   Reasoning along  similar lines,  we can simplify the expression for
the tensor $\mu_{ij}$, $\varphi_{ij}$, $\mu_i$,
which assumes the following form:

$$ \mu_{ij} = \sum_{bc} \frac {\chi^2 e_b e_c  m_b m_c n_b n_c c^3}
{16 \pi^3}\int \frac {d^4 p'}{\sqrt {(-g)}}\int
\frac {d^4   p''}{\sqrt   {(-g)}}{\lbrack}(z^2+\frac{1}{2})(   u''_{i}
u''_{j}+ $$
$$ + u'_i u'_j)g^{qr} + (z^2 - \frac {1}{2})g_{ij}g^{qr}  -  2z  (u'_i
u''_j + u''_i u'_j) g^{qr} - (z^2 - \frac {1}{2})
\left( \delta^{q}_{i}      \delta^{r}_{j}       +       \delta^{q}_{j}
\delta^{r}_{i}\right){\rbrack}\times$$
     \beq \label {87}
\times \left( z \delta^{m}_{f} - u''_f u'^m \right)J_{rqm}(u',  u'') f_c(x'')
\frac {\pa f_b(x')}{\pa p'_f},
     \eeq

$$\varphi^{ik} = \sum_{bc} \frac {\chi e_b e^2_c  n_b n_c
c }{2\pi}\int \frac {d^4 p'}{\sqrt {(-g)}}\int
\frac {d^4 p''}{\sqrt {(-g)}}(u'u'')K_{fl}(u', u'') \times $$
   \beq \label {88}
\times{\lbrack}(u'u'')(u'^i g^{kf}-u'^k g^{if}) -
(u''^i g^{kf}-u''^k g^{if}){\rbrack}
\left( f_c(x'') \frac {\pa f_b(x')}{\pa p'_{l}} -
f_b(x') \frac {\pa f_c(x'')}{\pa p''_{l}}\right),
     \eeq

$$ \mu^{i} = \sum_{bc} \frac {\chi e^2_b e_c  m_c n_b n_c c}{4\pi}
\int \frac {d^4 p'}{\sqrt {(-g)}}\int
\frac {d^4 p''}{\sqrt {(-g)}}{\lbrack}((u'u'')\delta^s_l - u''_l u'^s)
{\rbrack}\times$$
     \beq \label {89}
u''^k J^i_{ks}(u',  u'') f_c(x'')
\frac {\pa f_b(x')}{\pa p'_l}
\eeq

Here $z = (u'u'')$.

In (\ref {87}), (\ref {89}) we have introduced the notation
$ J_{rqm}(u',  u'') $ for tensor that in locally Lorentzian  reference
frame have the form
$$ J_{lmn}(u',u'') = \frac{1}{u'^0 u''^0}
\int\frac{d^3 k}{k^3}\int_{-\infty}^{\eta}   d\eta' \int_{-\infty}^{\eta}
d\eta''\int_{-\infty}^{\eta'}    d\tau' \int_{-\infty}^{\tau'}
d\tau'' {\bigl(} k^{+}_{l}e^{-ik(\eta'-\eta)} - $$
 $$ -k^{-}_{l}e^{ik(\eta'-\eta)}{\bigr)}
{\bigl(} k^{+}_{m}e^{ik(\eta''-\eta)} -
k^{-}_{m} e^{-ik(\eta''-\eta)}{\bigr)}
{\bigl(} k^{-}_{n}e^{ik(\tau''-\tau')} - $$
$$ -k^{+}_{n} e^{-ik(\tau''-\tau')}{\bigr)}exp[\frac {i}{c}
({\bf kv''})(\eta''-\tau'') +
     \frac{i}{c}({\bf kv'})({\tau}'-{\eta}')]. $$

     After evaluating  the  integrals  with  respect  $\eta',  \eta'',
\tau'$ and $ \tau'' $, we get
 $$ J_{lmn}(u',u'') =    \frac{c^4}{u'^0 u''^0}
\int \frac{d^3 k}{k^3} \frac {V.p.}{{\bf   (kv''-    kv')}}
{\biggl\{} \frac {k_{l}^{+}k_{m}^{+}k_{n}^{+}}{(kc+ {\bf kv''})^3} + $$
 $$ + \frac {k_{l}^{+}k_{m}^{+}k_{n}^{-} + k_{l}^{+}k_{m}^{-}k_{n}^{+} +
 k_{l}^{-}k_{m}^{+}k_{n}^{+}}{(kc+ {\bf kv''})^2(kc - {\bf kv''})} +
 \frac {k_{l}^{+}k_{m}^{-}k_{n}^{-} + k_{l}^{-}k_{m}^{+}k_{n}^{-} +
 k_{l}^{-}k_{m}^{-}k_{n}^{+}}{(kc+ {\bf kv''})(kc - {\bf kv''})^2} + $$
     \beq \label {89a}
+ \frac {k_{l}^{-}k_{m}^{-}k_{n}^{-}}{(kc - {\bf kv''})^3}{\biggr\}}
     \eeq

     The symbol $ V.p.$ indicates that the integral is calculated as a
principal value.

     Just as  in the previous case,  we specify (\ref {89a}) in
the center-of-mass reference frame, where
$${\bf v'} = {\bf v},\qquad {\bf v''} = -{\bf v},
\qquad u'^0 = u''^0 = 1/\sqrt {1- v^2/c^2} =u^0 $$.
In this reference frame the components of $ J_{lmn}(u',u'')$
have the following form (the  spatial  indexes  of  three-dimensional
velocity $ v^{\alpha}$ are lowered by using the tree-dimensional
Kronecker symbol
 $\delta_{\alpha \beta}$)
     \beq \label {89b}
 J_{000} = - \alpha (v) \frac {v^2}{c^2}, J_{00\alpha} = - \alpha (v)
 \frac {v_{\alpha}}{c},  \qquad  J_{0\alpha \beta} = - \alpha   (v)
\delta_{\alpha \beta} + \beta (v) \left( \delta_{\alpha \beta} -
\frac {v_{\alpha} v_{\beta}}{v^2}\right),
     \eeq
 $$ J_{\alpha \beta \gamma} = -\frac {c^2}{v^2}\alpha (v)
{\Bigl[} \delta_{\alpha \beta} \frac {v_{\gamma}}{c} +
 \delta_{\alpha \gamma} \frac {v_{\beta}}{c} +
\delta_{\beta \gamma}         \frac         {v_{\alpha}}{c}    -
2\frac{v_{\alpha}v_{\beta}v_{\gamma}}{c v^2}{\Bigr]} + $$
     \beq \label {89c}
+\frac {c^2}{v^2}\beta (v){\Bigl[} \left( \delta_{\alpha \beta} -
\frac {v_{\alpha}v_{\beta}}{v^2}\right)\frac {v_{\gamma}}{c} +
\left( \delta_{\alpha \gamma} - \frac {v_{\alpha}v_{\gamma}}{v^2}\right)
\frac {v_{\beta}}{c} +
\left( \delta_{\beta \gamma} - \frac {v_{\beta}v_{\gamma}}{v^2}\right)
\frac {v_{\alpha}}{c}{\Bigr]}.
     \eeq

    The function $\alpha $ and $\beta $ in (\ref {89b}) and (\ref {89c})
depend  on the velocity
$ v  =  \sqrt  {v_{1}^{2}  +  v_{2}^{2}  +  v_{3}^{2}}  $ only and
have the explicit form
     \beq \label {89d}
\alpha = \frac {\pi c^3}{u_{0}^{2} v^3 k_{min}}{\biggl[} \frac {2\frac {v}{c}
\left( 1       +       \frac       {v^2}{c^2}\right)}{\left(        1-
\frac{v^2}{c^2}\right)^2} + \ln \left( \frac {1- \frac {v}{c}}
{1 + \frac {v}{c}}\right){\biggr]},
     \eeq
     \beq \label {89e}
\beta = \frac {\pi c^3}{2 u_{0}^{2} v^3 k_{min}}{\biggl[} \frac {2\frac {v}{c}
\left( 3 - 2\frac {v^2}{c^2} + 3\frac {v^4}{c^4}\right)}{\left(        1-
\frac{v^2}{c^2}\right)^2} + 3\left( 1 + \frac {v^2}{c^2}\right)
 \ln \left( \frac {1- \frac {v}{c}}
{1 + \frac {v}{c}}\right){\biggr]}.
 \eeq
Here we introduced the following notation for the integral
     $$ \frac {1}{k_{min}} = \int_{k_{min}}^{\infty} \frac {dk}{k^2}. $$

     We set the lower
integration limit to $k_{min} =  1/r_{D}$.

     A covariant generalization of this results,  which were  obtained
in the locally Lorentzian center-of-mass reference frame, to arbitrary
reference frames has the form
 $$ J_{ijk}(u',u'') = A {\Bigl[} (g_{ij}u'_k +g_{ik}u'_j + g_{jk}u'_i) -
z(g_{ij} u''_k +g_{ik} u''_j + g_{jk} u''_i) - $$
$$ - (u'_i u''_j u''_k + u''_i u'_j u''_k + u''_i u''_j u'_k) +
3z u''_i u''_j u''_k {\Bigr]} + $$
$$ + C {\Bigl[} u'_i u'_j u'_k - z (u'_i u'_j u''_k + u'_i u''_j u'_k +
u''_i u'_j u'_k) + $$
     \beq \label {91}
 + z^2 (u'_i u''_j u''_k + u''_i u'_j u''_k + u''_i u''_j u'_k) -
 z^3 u''_i u''_j u''_k {\Bigr]},
     \eeq
     where $ z = (u' u'') = (u'^i u''_i) $ ,
     \beq \label {92}
 A  =  -  \frac  {2\pi  \sqrt  {2}}{k_{min}}{\biggl[} \frac {(z-2)}{(z-1)^2
(z+1)^{1/2}} + \frac {(2z-1)}{(z+1)(z-1)^{5/2}} \ln \left( z +
\sqrt {z^2 - 1}\right) {\biggr]},
     \eeq
     \beq \label {93}
 C = - \frac {2\pi \sqrt {2}}{k_{min}}{\biggl[} \frac {(z-6)}{(z-1)^3
(z+1)^{3/2}} + \frac {(6z-1)}{(z+1)^2 (z-1)^{7/2}} \ln \left( z +
\sqrt {z^2 - 1}\right) {\biggr]}.
     \eeq

     The tensor $ J_{ijk}(u',u'') $ satisfies the identity
     \beq \label {95}
     J_{ijk}(u',u'') u''^{k} = 0.
     \eeq

Note that the tensor $\mu_{ij}$ is traceless:
     \beq \label {97}
     g^{ij} \mu_{ij} = 0.
     \eeq

     Let us now simplify the tensor  $T_{ij}^{(r)}$
(see. (\ref {36}).  Substitution of (\ref {62}) to (\ref {36})  yields
the expression for  $T_{ij}^{(r)}$. In view  (\ref
{70}) - (\ref {73a}) we have  (if $n_a \gg 1$):

$$ T^{(r)}_{ij} = \frac {1}{4\pi}\left(-g_{it}g_{js} + \frac {1}{4} g_{ij}
g_{ts}\right)\sum_{bc}\int d^4  p'\int  d^4p''\int  d^3  q'  \int  d^3
q''\times $$
$$\times \int_{-\infty}^{\eta}   d {\eta}'
\int_{-\infty}^{\eta}d{{\eta}''}\int d^3 k' \int d^3 k''
e^{-i{\bf k'}({\bf q}-{\bf  q'})}
e^{-i{\bf      k''}({\bf      q}-{\bf     q}'')}\times$$
     \beq \label {98}
\times \omega^{t(b)}_{.l}(\eta,{\eta}', p',{\bf k'})\omega^{(c)sl}
(\eta,{\eta}'',p'',{\bf k''})  n_b  n_c  g_{bc}(x',x''),
\eeq

Here we can not  neglect by $g_{bc}^{(gr)}(x',  x'')$ in  expression
(\ref {76}) for $g_{bc}(x', x'')$.

 That is why
     \beq \label {99}
    T_{ij}^{(r)} = \tau_{ij}^{(r)} + \tau_{ij}^{(gr)}.
     \eeq

One can get  $\tau_{ij}^{(r)}$ if replace $g_{bc}(x',  x'')$ by
$g_{bc}^{(el)}(x',  x'')$  (see (\ref {79})) in expression  (\ref {98})
and replace $g_{bc}(x',  x'')$ by $g_{bc}^{(gr)}(x', x'')$
(see (\ref {77}) to obtain the expression for $\tau_{ij}^{(gr)}$.

Substitution (\ref {77}) and (\ref {79}) to  (\ref  {98})  yields  the
folloing expressions for $\tau_{ij}^{(r)}$ and $\tau_{ij}^{(gr)}$:

$$ \tau_{ij}^{(r)} = \sum_{bc} \frac {e^2_b e^2_c  n_b n_c}
{2 (2\pi)^4 c}\int \frac {d^4 p'}{\sqrt {(-g)}}\int
\frac {d^4   p''}{\sqrt   {(-g)}}{\lbrack}-2 z \delta_i^r   \delta_j^f  +
z g_{ij}g^{rf} + $$
$$+ (\delta_i^f  {u''}_j + \delta_j^f {u''}_i ) {u'}^r - g^{rf}({u'}_i
{u''}_j + {u''}_i {u'}_j){\rbrack}\times$$
     \beq \label {100}
\times \left( z \delta^{s}_{n} - u''_n u'^s \right)J^{(el)}_{rfs}(u',  u'')
f_c(x'')
\frac {\pa f_b(x')}{\pa p'_n}.
     \eeq
$$ \tau_{ij}^{(gr)} = \sum_{bc} \frac {\chi e_b e_c m_b m_c n_b n_c c^3}
{16 \pi^2}\int \frac {d^4 p'}{\sqrt {(-g)}}\int
\frac {d^4 p''}{\sqrt {(-g)}}{\lbrack} 2 z \delta_i^p \delta_j^q -
z g_{ij}g^{pq} - $$
$$- (\delta_i^q  {u''}_j + \delta_j^q {u''}_i ) {u'}^p + g^{pq}({u'}_i
{u''}_j + {u''}_i {u'}_j){\rbrack} J^{(gr)}_{pqf}(u',  u'') f_c (x'')\times$$
     \beq \label {101}
\times \frac  {\pa  }{\pa  p'_n}.\Big\{  f_b(x')
\big[\left(z^2-\frac {1}{2}\right) \delta_n^f +
\left(z^2+\frac {1}{2}\right){u'}_n   {u'}^f  -  2z  {u''}_n
{u'}^f\big]\Big\}.
     \eeq

Note that the tensors $\tau_{ij}^{(gr)}$ and $\tau_{ij}^{(r)}$ are traceless:
     \beq \label {101a}
     g^{ij} \tau^{(gr)}_{ij} = 0, \qquad g^{ij} \tau^{(r)}_{ij} = 0
     \eeq

In (\ref {100}) and (\ref {101}) the tensors $J^{(el)}_{rpq}(u', u'')$
and $J^{(gr)}_{rpq}(u',  u'')$ have the form (\ref {91}), where $A$ and
$B$ have the forms (\ref {92}) and (\ref {93}) respectively. But in the
expression for $J^{(el)}_{rpq}(u',  u'')$ we must  put  $k_{min}  =
1/r_D$, where $r_D$ is the radius of Debit,  since the electromagnetic
interaction in plasma are shielded under $r>r_D$. In the expression for
$J^{(gr)}_{rpq}(u', u'')$  we  must    put $k_{min} = 1/r_g$,  where
$r_g$ is the radius of correlation for gravitational interaction.
     As the experience of deriving the relativistic  kinetic  equation
     (refer to. [13],  [14],  [16], [20]) shows, more thorough investigations
suggest than the integrals become convergent as $r \rightarrow \infty $,
with the contribution from the region where $ r > r_g $ being
infinitesimal. In Ref. [14],[16] there are estimates for
$r_g$ in the case
where the  average metric $g_{ij}$ is the metric of isotropic
cosmological model and in the case of gravitational interaction of particles.

     \bigskip

{\bf 4.  Macroscopic system of Einstein and Maxwell equations for
relativistic plasma}

     As a result were obtained  the  macroscopic  Einstein  and  Maxwell
equations in relativistic plasma. They have the forms:
     \beq \label {102}
     G_{ij} +    \nabla_k    \varphi^k_{ij}    +   \mu_{ij}   -   \chi
\tau_{ij}^{(gr)} = \chi T_{ij},
     \eeq
     \beq \label {103}
     \nabla_k F^{ik}   +   \nabla_k  \varphi^{ik}  +  \mu^i  =  -\frac
{4\pi}{c} J^i.
     \eeq
     Here $G_{ij}$ is the Einstein's tensor of  the  Riamannian  space
with  macroscopic  metric  $g_{ij}$,  $F^{ik}$  is the macroscopic
tensor of electromagnetic field (Maxwell's tensor),
$J^i$ is macroscopic current vector,
$T_{ij}$ is the macroscopic energy-momentum tensor. The last is the sum  of
macroscopic energy-momentum  tensor  of medium $T_{ij}^{(m)}$
(\ref {34}),  energy
momentum tensor of macroscopic electromagnetical field $T_{ij}^{(el)}$
(\ref {29}) and macroscopic energy-momentum tensor
$\tau_{ij}^{(r)}$ (\ref   {100})  of  electromagnetical  radiation  in
plasma. (In cosmological plasma  in  the  last  case  one  should  say
about the energy - momentum thensor of relict radiation.)

The Einstein equations of the gravitational field for continium media,
obtained here,  differ from the classical Einstein equations  by
the presence of additional terms
     $\nabla_k \varphi^k_{ij}$, $\mu_{ij}$ and $- \chi \tau_{ij}^{(gr)}$
in the left-hand side. It caused by particle interaction.
The forms of this tensors are
(\ref  {86}),  (\ref {87}) and (\ref {101}).
The third  term,  $(-\chi \tau_{ij}^{(gr)})$ is the addition to macroscopic
energy - momentum  tensor  of  electromagnetic  radiation,  caused  by
gravitational interaction  which multiplying on $\chi$ and moving from
the right-hand side of macroscopic equations to the left-hand side.

The macroscopic Maxwell equations differ from  the  classical  Maxwell
equations by the presence of additional terms
$\nabla_k \varphi^{ki} + \mu^i$.  The additional terms in Maxwell
equations caused by particle interaction and  by  effects  of  general
relativity.

     The tensors
     $\nabla_k \varphi^k_{ij}$, $\mu_{ij}$, $ \tau_{ij}^{(gr)}$,
     $\nabla_k \varphi^{ki}$ and $\mu^i$ are expressed in
     (\ref {86})  ---  (\ref  {89})  and  (\ref  {101})  in  terms  of
one-particle distribution    function    $f_b$    specified   in   the
eight-dimensional phase spase in which all four components of momentum
are independent.  The transition to the seven-dimensional distribution
function  $F_a (q^i, p_{\alpha})$ is made according to the formula
     \beq \label {104}
     n_a f_a  (q^i,  p_j)  =  F_a   (q^i,   p_{\alpha})\delta   (\sqrt
{g^{lm}p_lp_m} - m_ac).
     \eeq
Here the  function  $F_a$  depends on  the spatial components of
momentum only. Greek indexes are used to denote spartial components.

By integrating (\ref {86}) ---  (\ref  {89})  and  (\ref  {101})  with
respect to   ${p'}_0$ and  ${p''}_0$ we can whrite down the tensors
     $\nabla_k \varphi^k_{ij}$, $\mu_{ij}$, $ \tau_{ij}^{(gr)}$,
     $\nabla_k \varphi^{ki}$ and $\mu^i$ as

$$\varphi^{k}_{ij} =
 \sum_{bc} \frac {\chi^{2} e_b e_c m^2_{b} m^2_{c} c^5 }{16( \pi)^2}
\int \frac {d^3 p'}{{p'}^0\sqrt {(-g)}}\int
\frac {d^3 p''}{{p''}^0\sqrt {(-g)}}{\biggl[}
\frac{1}{2} g^{fk} u''_{i} u''_{j} + $$
     \beq \label {105}
 +u'^{k}  (u'  u'')(\delta^{f}_{j}  u''_{i}  +  \delta^{f}_{i}
u''_{j}){\biggr]} (u' u'')  K_{f\alpha}(u',u'')
\left( F_c(x'') \frac {\pa F_b(x')}{\pa p'_{\alpha}} -
F_b(x') \frac {\pa F_c(x'')}{\pa p''_{\alpha}}\right)
     \eeq
$$ \mu_{ij} = \sum_{bc} \frac {\chi^2 e_b e_c  m^2_b m^2_c c^5}
{16 \pi^2}\int \frac {d^3 p'}{{p'}^0\sqrt {(-g)}}\int
\frac {d^3   p''}{{p''}^0\sqrt   {(-g)}}{\lbrack}(z^2+\frac{1}{2})(   u''_{i}
u''_{j}+ $$
$$ + u'_i u'_j)g^{qr} + (z^2 - \frac {1}{2})g_{ij}g^{qr}  -  2z  (u'_i
u''_j + u''_i u'_j) g^{qr} - (z^2 - \frac {1}{2})
\left( \delta^{q}_{i}      \delta^{r}_{j}       +       \delta^{q}_{j}
\delta^{r}_{i}\right){\rbrack}\times$$
     \beq \label {106}
\times \left( z \delta^{m}_{\alpha} - u''_{\alpha} u'^m \right)
J_{rqm}(u',  u'') F_c(x'')
\frac {\pa F_b(x')}{\pa p'_{\alpha}}.
     \eeq

$$\varphi^{ik} = \sum_{bc} \frac {\chi e_b e^2_c m_b m_c c^3 }{2\pi}
\int \frac {d^3 p'}{{p'}^0\sqrt {(-g)}}\int
\frac {d^3 p''}{{p''}^0\sqrt {(-g)}}(u'u'')K_{f \alpha}(u', u'') \times $$
   \beq \label {107}
\times{\lbrack}(u'u'')(u'^i g^{kf}-u'^k g^{if}) -
(u''^i g^{kf}-u''^k g^{if}){\rbrack}
\left( F_c(x'') \frac {\pa F_b(x')}{\pa p'_{\alpha}} -
F_b(x') \frac {\pa F_c(x'')}{\pa p''_{\alpha}}\right),
     \eeq

$$ \mu^{i} = \sum_{bc} \frac {\chi e^2_b e_c  m_b m_c^2 c^3}{4\pi}
\int \frac {d^3 p'}{{p'}^0\sqrt {(-g)}}\int
\frac {d^3 p''}{{p''}^0\sqrt {(-g)}}{\lbrack}((u'u'')\delta^s_{\alpha} -
u''_{\alpha} u'^s){\rbrack}\times$$
     \beq \label {108}
u''^k J^i_{ks}(u',  u'') F_c(x'')
\frac {\pa F_b(x')}{\pa p'_{\alpha}}
\eeq

$$ \tau_{ij}^{(gr)} = \sum_{bc} \frac {\chi e_b e_c m^2_b m^2_c c^5}
{16 \pi^2}\int \frac {d^3 p'}{{p'}^0\sqrt {(-g)}}\int
\frac {d^3 p''}{{p''}^0\sqrt {(-g)}}{\lbrack} 2 z \delta_i^p \delta_j^q -
z g_{ij}g^{pq} - $$
$$- (\delta_i^q  {u''}_j + \delta_j^q {u''}_i ) {u'}^p + g^{pq}({u'}_i
{u''}_j + {u''}_i {u'}_j){\rbrack} J^{(gr)}_{pqf}(u',  u'') F_c (x'')\times$$
     \beq \label {109}
\times \frac  {\pa  }{\pa  p'_n}.\Big\{  F_b(x')
\big[\left(z^2-\frac {1}{2}\right) \delta_n^f +
\left(z^2+\frac {1}{2}\right){u'}_n   {u'}^f  -  2z  {u''}_n
{u'}^f\big]\Big\}.
     \eeq

     Here
$$\frac  {d^3  p'}{p'^0  \sqrt  {(-g)}} \hspace{1cm}
and \hspace{1cm}\frac  {d^3  p''}{p''^0  \sqrt  {(-g)}}$$
are the invariant volume elements in tree - dimensional momentum spase
of particles spesies "b" and "c" respectively.

The greek index $\alpha$ in (\ref {105}) - (\ref {108})  takes
the values 1,2 and 3 only (the spartial index).  The derivative with respect
to $p'_n$ in (\ref {109}) should by calculated as all four  components
of momentum are independent. The dependence of $p'_0$ on $p'_{\alpha}$
is taken into account after differentiation with  respect  $p'_n$
is completed only.

In (\ref {108}) and (\ref {109}) the tensors $J^{(el)}_{rpq}(u', u'')$
and $J^{(gr)}_{rpq}(u',  u'')$ have the form (\ref {91}), where $A$ and
$B$ have the forms (\ref {92}) and (\ref {93}) respectively. But in
expression for $J^{(el)}_{rpq}(u',  u'')$ we must   put  $k_{min}  =
1/r_D$, where $r_D$ is the radius of Debit,  since the electromagnetic
interaction in plasma are shielded under $r>r_D$. In the expression for
$J^{(gr)}_{rpq}(u', u'')$  we  must   put $k_{min} = 1/r_g$,  where
$r_g$ is the radius of correlation for gravitational interaction.
     As the experience of deriving the relativistic  kinetic  equation
     (refer to. [13],  [14],  [16], [20]) shows, more thorough investigations
suggest than the integrals become convergent as $r \rightarrow \infty $,
with the contribution from the region where $ r > r_g $ being
infinitesimal. In Ref. [14],[16] there are estimates for
$r_g$ in the case
where the  average metric $g_{ij}$ is the metric of isotropic
cosmological model and in the case of gravitational interaction of particles.

     The tensors $  \varphi^{k}_{ij}  $ , $ \mu_{ij} $, $\tau_{ij}^{(gr)}$
и $\mu^i$  must obey the additional conditions
     \beq \label {110}
  g^{lj}\nabla_l \left( \nabla_k \varphi^{k}_{ij} + \mu_{ij}  -
\chi \tau_{ij}^{(gr)}\right) = 0,
     \eeq
     \beq \label {111}
     \nabla_i \mu^i = 0,
     \eeq
since the   divergenses  of  $G_{ij}$,  $T_{ij}$,  $\nabla_k  F^{ik}$,
$\nabla_k \varphi^{ik}$, $J^i$ vanish.

Equations (\ref {110}),  (\ref {111}) impose some restrictions  on
the  parametres $r_D$ and $r_g$  dependence  on the coordinates
and the relative velosity of particles. The latter can be expressed via
of $z=(u'u'')$.

The macroscopic  energy-momentum  tensor  $T_{ij}^{(m)}$ of medium and
the current  vector  $J^i$  can  by  alsow  written  in   terms   of
one-particle distribution function as follows:
     \beq \label {112}
     T_{ij}^{(m)} = \sum_{a} c \int \frac {d^3 p}{p^0 \sqrt {(-g)}} p_i p_j
F_a(p),
     \eeq
     \beq \label {113}
     J^i =  \sum_{a} e_a c \int  \frac  {d^3  p}{p^0 \sqrt {(-g)}} p^i F_a(p).
     \eeq

The system  of equations (\ref {102}).  (\ref {103}) must by augmented
by the kinetic equation for $F_b$ in relativistic plasma.  In the case
when the  electromagnetical  interaction  og particles are dominating,
the equation for $F_a$ was derived in Refs. [11], [13].

The covariant form of this kinetic equations is
   $$  u^{i}\frac{\pa   f_a}{\pa  q^i}  +  \Gamma_{j,ik}\frac  {\pa
f_a}{\pa p_i} + \frac {e_a}{c}<F_{ik}>u^k \frac{\pa f_a}{\pa p_i} = $$
     \beq \label {114}
=\sum_{b}  \frac {\pa}{\pa p_i}   \int     \frac     {d^4p'}{\sqrt
{(-g)}} E_{ij}(p, p') \left( \frac {\pa  f_a}{\pa  p_j} f'_{b} -
\frac {\pa
f'_b}{\pa p'_j}f_{a}\right),
     \eeq

     where
   $$  E_{ij}(p,p')  = \frac {2\pi e_{a}^{2} e_{b}^{2} L n_b}{c^2}[(u,u')^2
- 1]^{-3/2}(u', u)^2\times  $$
     \beq \label {115}
  \times {\bigl\{} -g_{ij}[(u,u')^2  -1]  -  u_i u_j  -   u'_iu'_j +
(u,u')(u_iu'_j +  u'_iu_j){\bigr\}}
     \eeq

     with $(u,u') =  u'_iu^i$ .
Primed and  nonprimed  quantities  refer  to  particles  belonging  to
species $a$, and $b$ respectively  and  $L$
is  the  Coulomb logarithm  [15]

     \beq \label {116}
     L = \int^{k_\infty}_{k_{min}}\frac {dk}{k}
     \eeq

\bigskip

{\bf 5. Conclusion}

     The macroscopic equations of the gravitational field in
relativistic plasma differ from the classical  Einstein  equations  by
the presence of additional terms
     $$Z_{ij} =  \nabla_k    \varphi^k_{ij}    +   \mu_{ij}   -   \chi
\tau_{ij}^{(gr)}$$
on the left-hand side due to partial interaction.

     These terms are proportional to  the  square  of  Einstein
constant and to the square of particle namber density.

     The macroscopic equations of  the  electromagnetical  field  in
relativistic plasma  differ  from  classical  Maxwell equations by the
presence of additional terms
     $$Z^i = \nabla_k \varphi^{ki} + \mu^i.$$
on the left-hand side due to particle interaction and due  to  effects
of general relativity.

     This terms  are  proportional  to  the  first  power  of Einstein
constant and to the square of the particle number density.

     Hence these terms can play an important role  in  continuous
media of  very  high  density only.  Such density are possible in the early
stages of the evolution of the Universe and inside  objects hear
gravitational collapse.   Therefore,   it   is  natural  to  look  for
applications of the derived equations primarily in the theory of early
stages of  the Universe evolution  and in gravitational collapse
theory.

\newpage
\pagestyle{empty}

\end{document}